\documentclass{aa}

\usepackage{txfonts}

\usepackage{graphicx}
\usepackage{color}
\usepackage{url}

\usepackage[]{natbib}
\bibpunct{(}{)}{;}{a}{}{,} 

\newcommand{\myfigure}[3]{
  \begin{figure}
    \resizebox{\hsize}{!}{\includegraphics{#1}}
    \caption{#2}
    \label{#3}
  \end{figure}
}

\begin{document}

\title{Exploring high-z galaxies with the E-ELT} 

\date{\today} 
\author{
  M. Gullieuszik\inst{1}\and
  R. Falomo\inst{1}\and
L. Greggio\inst{1}\and
 M. Uslenghi\inst{2} \and
 D. Fantinel\inst{1}
}
\institute{
  INAF, Osservatorio Astronomico di Padova, 
  Vicolo dell'Osservatorio 5, I-35122 Padova, Italy
  \and
 INAF, Istituto di Astrofisica Spaziale e Fisica Cosmica, 
 Via Bassini 15, I-20133 Milano, Italy 
  }

\abstract{ We present simulated observations of galaxies
  at $z=2$ and $z=3$ to probe the capabilities of next-generation
  telescopes (E-ELT and JWST) to measure the structural and
  photometrical properties of high-redshift galaxies.  We carry out an
  extensive set of simulations of high-redshift galaxies adopting the
  specifications of the E-ELT first light instrument MICADO.  The main
  parameters (sizes, Sersic index, and magnitudes) of the galaxies are
  measured using GALFIT and the accuracy of the determinations is
  assessed by comparing the input values to the measurements from many
  runs with different statistical noise.  We also address the effects
  on the accuracy of the measurements of possible spatial variation of the
  point spread function (PSF) in the field.  
  We find that from $3 h$ exposure E-ELT near-infrared (IR)
  images of galaxies at $z \sim 2$ and $z \sim 3$ it will be possible
  to measure the size, total magnitude, and galaxy morphology
  with an accuracy of 2-5\% for objects as faint as $H\sim$ 25 and
  half-light size of 0.2 arcsec.  The effective radius of compact,
  early-type galaxies is also recovered with $\sim 5$\% accuracy,
  provided that their half-light size exceeds 20 mas.  These results
  are compared with those expected from simulated observations
  obtained with NIRCam on board the JWST.}

\maketitle

\section{Introduction}\label{sec:intro}

Understanding the assembly history of galaxies is of paramount importance for answering fundamental questions about the processes of
formation and evolution of galaxies and their associations (groups, clusters, superclusters, etc). 
To this end, it is of extreme relevance to be able to characterise the properties of galaxies at
high redshift  to probe their evolution over a significant interval of cosmic time. 
Their global structure and colours yield insight into the conditions of star formation and the subsequent merger events and/or secular processes
\citep[e.g.][]{dalcanton+1997,mo+1998}.

Although galaxies usually comprise multiple different components
(i.e. bulges, disks, substructures) fitting their surface brightness
profile with  a single Sersic law can provide relevant basic data
\citep[e.g.][]{kelvin+2012}.  This is in particular needed for the
study of distant galaxies that are poorly resolved with current ground-based telescopes and partially available with Hubble Space Telescope
\cite[HST; e.g.][]{vand+2012}.

Unfortunately the characterisation of photometrical and structural
properties of high-$z$ objects is hampered by the faintness of the
targets and by their very small angular size.
Hubble Space Telescope observations have shown
that galaxies at high-$z$ are significantly smaller in size than
galaxies of similar mass at low $z$
\citep[e.g.][]{dadd+2005,truj+2006,buit+2008,vulcani+2014,kennedy+2015}
and this size evolution ($R_e\sim (1+z)^{-1}$) with redshift, furthermore, hinders
 the capability of studying the structural properties of
these galaxies.  In fact present estimates of galaxy size at
$z\gtrsim2$ yield apparent sizes of $\sim 100-200$ mas or less. At
higher redshift the situation is even more challenging since the
angular size of the galaxies become smaller than 50 mas
\citep{ono+2013}.
As a consequence, it is now possible to  characterise these galaxies only via HST observations, but this is limited to galaxies with
mass $\gtrsim10^{10}M_\sun$\citep[][and refs
  therein]{bram+2012,vand+2012,vand+2014}.  In spite of the
diffraction limit images obtained by HST, the size of these high-redshift galaxies is so small that a significantly better spatial
resolution is needed to be able to properly characterise their
structure.
A major improvement in this direction is expected by future
observations gathered by the James Webb Space Telescope (JWST), which also has the important advantage of an extremely low background in the
near-infrared (IR). Because of the relative small aperture (6.5m), however, the JWST
resolution is limited to few hundredths of arcsec.  A more significant
step is expected by the future generation of extremely large ground-based optical/near-IR telescopes.  In fact these 30-40 m aperture
telescopes assisted by adaptive optics can reach resolution of few mas
in the near-IR.

In this paper we aim to address the imaging capabilities of the planned future 
instrumentation for Extremely Large Telescopes (ELTs) using the expected performances of the Multi-Adaptive Optics 
Imaging Camera for Deep Observations \citep[MICADO;][]{davi+2010}   at the European Extremely Large Telescope (E-ELT)\footnote{\url{http://www.eso.org/sci/facilities/eelt/}}
as a reference. We quantify the accuracy with which it will be possible to characterise the properties of 
high-redshift galaxies with  simulated MICADO observations.  In particular these simulations are used to investigate the possibility to 
accurately measure both the size and morphology of the galaxies. Moreover  using multi-band observations we  estimate  the accuracy 
that could be achieved in the measurements of the colour gradient; this is a key tool to test models of galaxies formation.
In fact the colour gradients trace variations of the properties of stellar populations, and are linked to the star
formation history of galaxies \citep[see e.g][]{sagl+2000,tamu+2000,koba+2004,garg+2011}.

Finally, the results obtained with simulated E-ELT observations are compared with those expected for the same targets from observations
secured with the Near-Infrared  camera (NIRcam) on board  the JWST\footnote{\url{http://www.stsci.edu/jwst/instruments/nircam}}.
Throughout this paper, we adopt a concordance cosmology with $H_0 = 70$ km s$^{-1}$ Mpc$^{-1}$, 
$\Omega_m = 0.3$, and $\Omega_\Lambda= 0.7$. All magnitudes are in AB system.

\section{Simulations of high-z galaxies}

In order to assess the accuracy with which it will be possible to characterise the
galaxy properties from observations with ELTs, we analyse simulated images  of high-redshift
galaxies assuming  a wide range of photometric and structural parameters. 
The comparison between the observed photometric and structural parameters and the true (input) values yields insights on the 
statistical and systematic effects. 

\subsection{Input models}  

In this work we assume simple smooth models for the simulation of galaxies. In particular Sersic models with various different parameters are chosen to 
study the effects for a variety of cases. More complex models (e.g.  substructures, knots, and tails) are not  considered, as
our main objective is to investigate the capability to derive the main global parameters (luminosity, scale-length, and morphology) of high-redshift galaxies.  

The Sersic law is defined by the following expression:
\begin{equation}
\mu(r)=\mu_e+2.5\frac{b_n}{\ln(10)}[(r/R_e)^{1/n}-1]
\end{equation}
\begin{equation}
\mu_e=M_{tot}+2.5\log[(1-e) \, 2\pi R_e^2 ]+
2.5\log[n e^{b_n} \, \Gamma(2n)/b_n^{2n}],
\end{equation}
where $M_{tot}$ is the total magnitude, $e$ the ellipticity,
$\Gamma$ the complete gamma function, and
the value of $b_n$ is such that
\begin{equation}
  2 \gamma(2n, \, b_n)=\Gamma(2n),
\end{equation}
where $\gamma$ is the incomplete gamma function
\citep[for details see][]{ciot1991}.

ELTs combined with adaptive optics systems will deliver the best
performances in the near-IR. Therefore we performed two sets of
simulations in the $J$, $H$, and $K$ bands to map the $U-V$ rest-frame
colour of galaxies at $z\sim2.3$ ($J$ and $H$ bands) and at $z\sim 3.3$
($H$ and $K$ bands). In the following, we refer to these two sets
of simulations as $z=2$ and $z=3$, respectively.  To test the
dependence of the results on the galaxy size and morphology, we
considered 27 template galaxies with nine values of the total mass in
the range $9<\log M/M_\sun<11$ and three values of the Sersic index,
$n=1.0$, 2.5, and 4.0.

\begin{figure*}
  \centering
  \includegraphics[width=8.3cm]{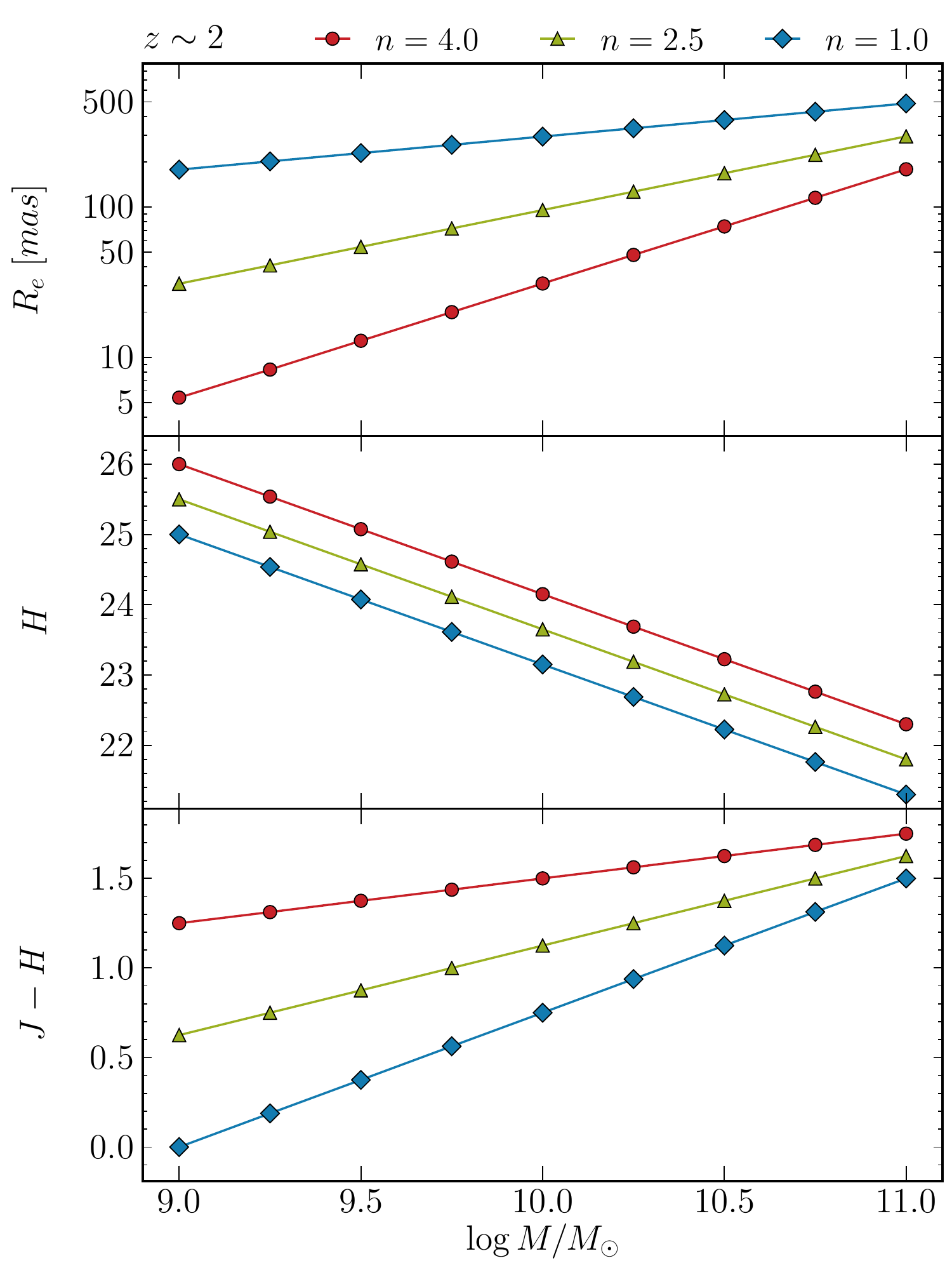}\hfill
  \includegraphics[width=8.3cm]{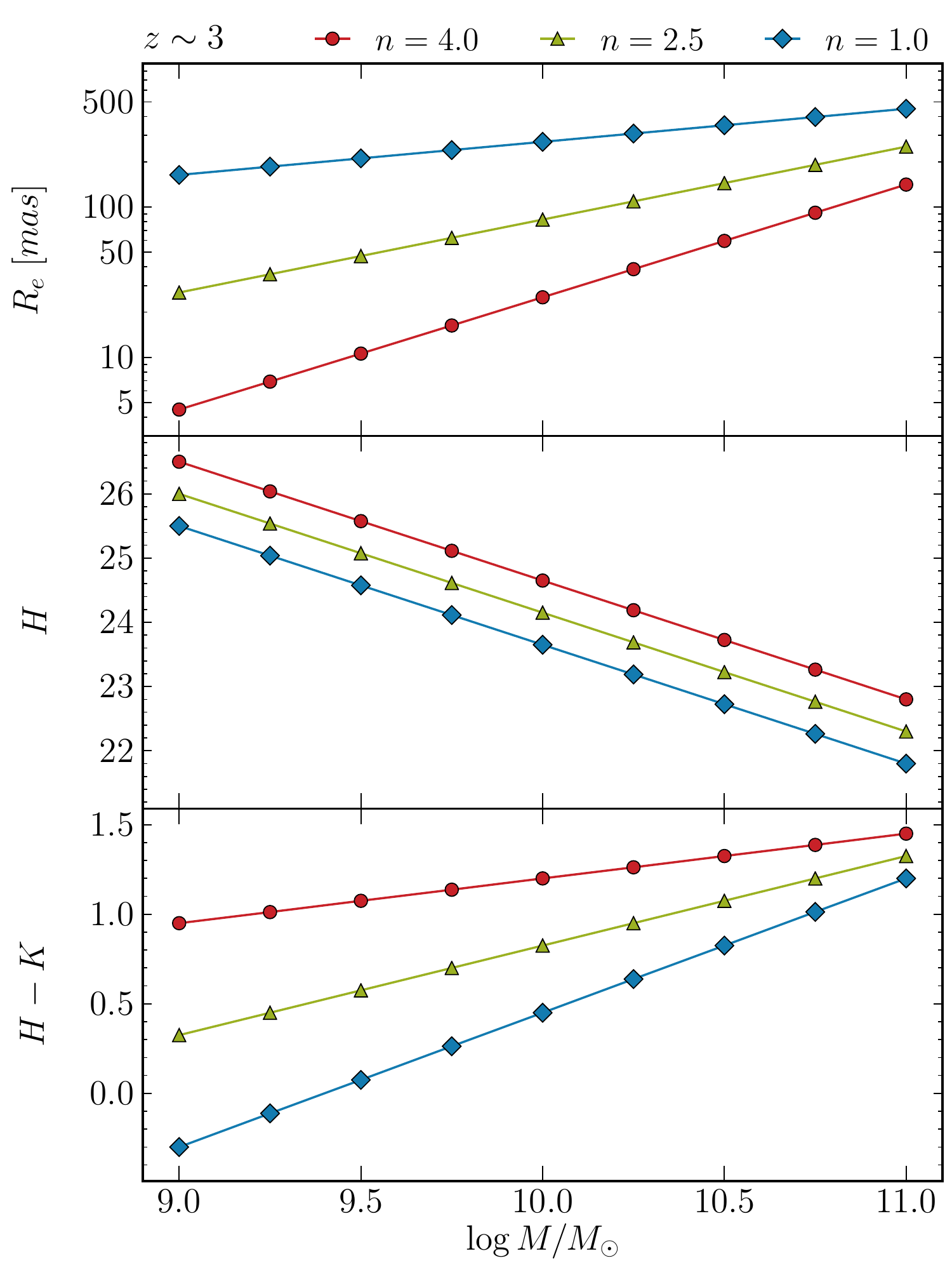}
  \caption{Structural and photometric parameters adopted for the two
    sets of simulated galaxies at $z=2$ ({\it left panels}) and $z=3$
    ({\it right panels}).  For each set we plot the relation between
    the galaxy mass and effective radii in milli-arcseconds ({\it
      top panel}), magnitude ({\it middle panel}), and colour ({\it
      bottom panel}) for the 9 template galaxies. In each panel we
    show the scaling relations for galaxies with 3 different
    Sersic index with different colours and symbols (see the legend on
    top of the figure).  }
  \label{fig:scalerel}
\end{figure*}

The relationships among the structural parameters ($R_e$ and $\mu_e$) and the galaxy masses for galaxies of the different morphological
types were calibrated following the results of \cite{vand+2014}, who
presented measurements for a sample of 30000 galaxies from the CANDELS HST survey with redshift $0<z<3$.  For simulations of $z=2$ galaxies,
we used the mean loci of $z=2.25$ galaxies.  For simulations of $n=1.0$ and $n=4.0$ galaxies, we used the scaling
relations for late-type and early-type galaxies, respectively.  For $n=2.5$ galaxies, we adopted the intermediate (average) values of above
relationships.  
The parameters for the simulated galaxies at $z=3$ were defined by extrapolating the scaling relations of
\cite{vand+2014} at redshift z=3.3. We used an ellipticity of 0.3 and a 90$\degr$ position angle (major axis along
the horizontal direction) for all galaxies. The complete list of adopted parameters for the simulated galaxies are shown in Fig. \ref{fig:scalerel} and
reported in Table \ref{tab:scalerelz2} and \ref{tab:scalerelz3} (available only in the online version of this paper).

\subsection{Image simulations}\label{sec:imasims}

Simulated images of galaxies were produced using the Advanced Exposure Time Calculator (AETC)\footnote{\url{http://aetc.oapd.inaf.it/}}  tool
\citep{falo+2011}. We adopt as baseline the parameters of MICADO, the instrument designed for the 39-m aperture E-ELT to provide quasi-diffraction limited
imaging over a wide ($\sim 1\arcmin$) field of view. The simulations
of MICADO observations were performed following the same prescription as in \cite{gull+2014}.
We used  the point spread functions (PSFs) of the multi-conjugate adaptive optics post focal relay \citep[MAORY;][]{diol+2010}
for the E-ELT, calculated for a $0\farcs6$ seeing at the centre of the corrected field of view
(FoV). These were produced for the final design of MAORY phase A study. 

The final simulated images were obtained as the stack of 180 individual images of 60s exposure for a total exposure time of three hours. 
To evaluate the effect of the statistical noise, we performed 50 runs for each simulated  galaxy and then compared the resulting parameters for all the simulations.

\begin{figure*}
  \centering
  \includegraphics[width=14.7cm]{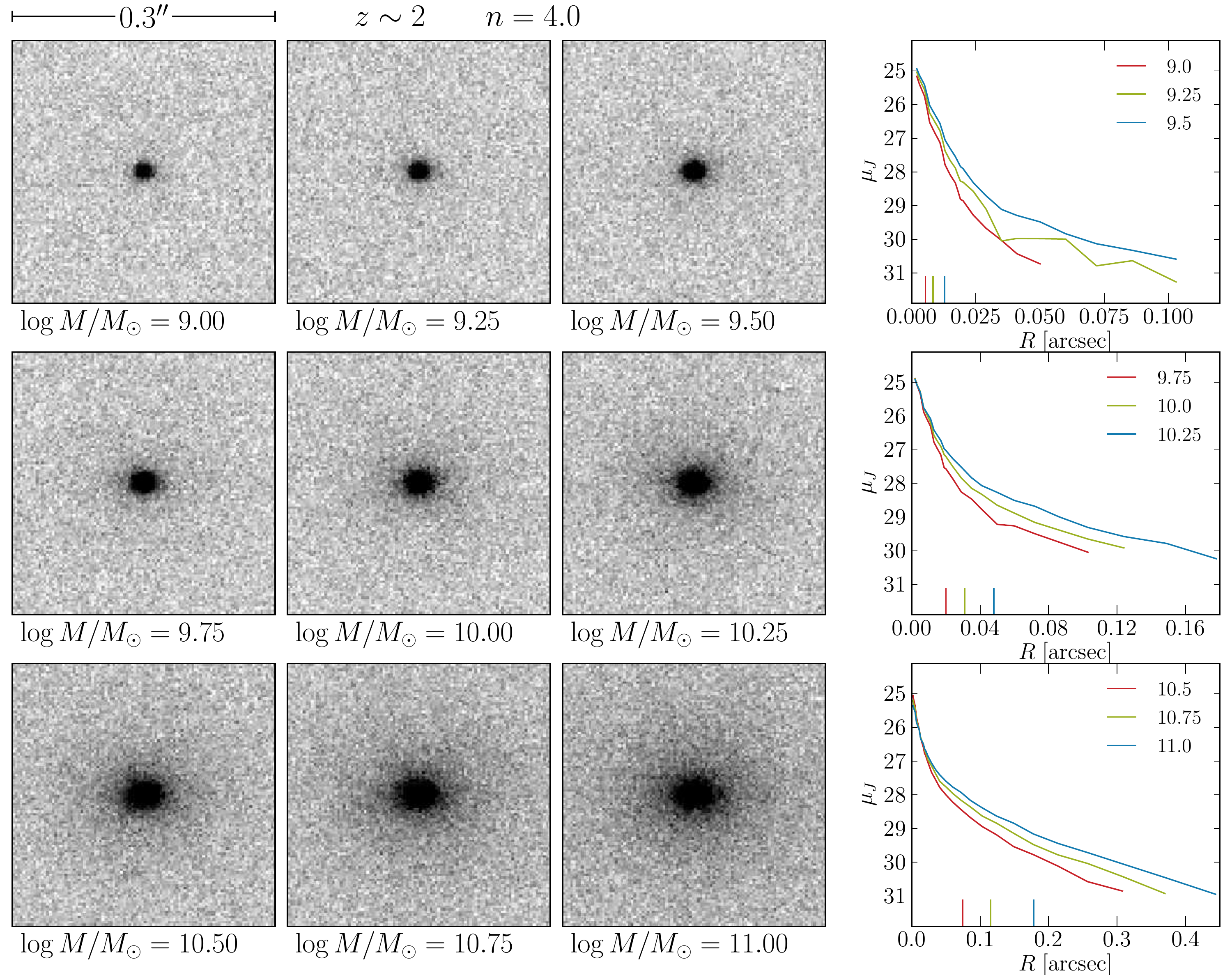}\\[1em]
  \includegraphics[width=14.7cm]{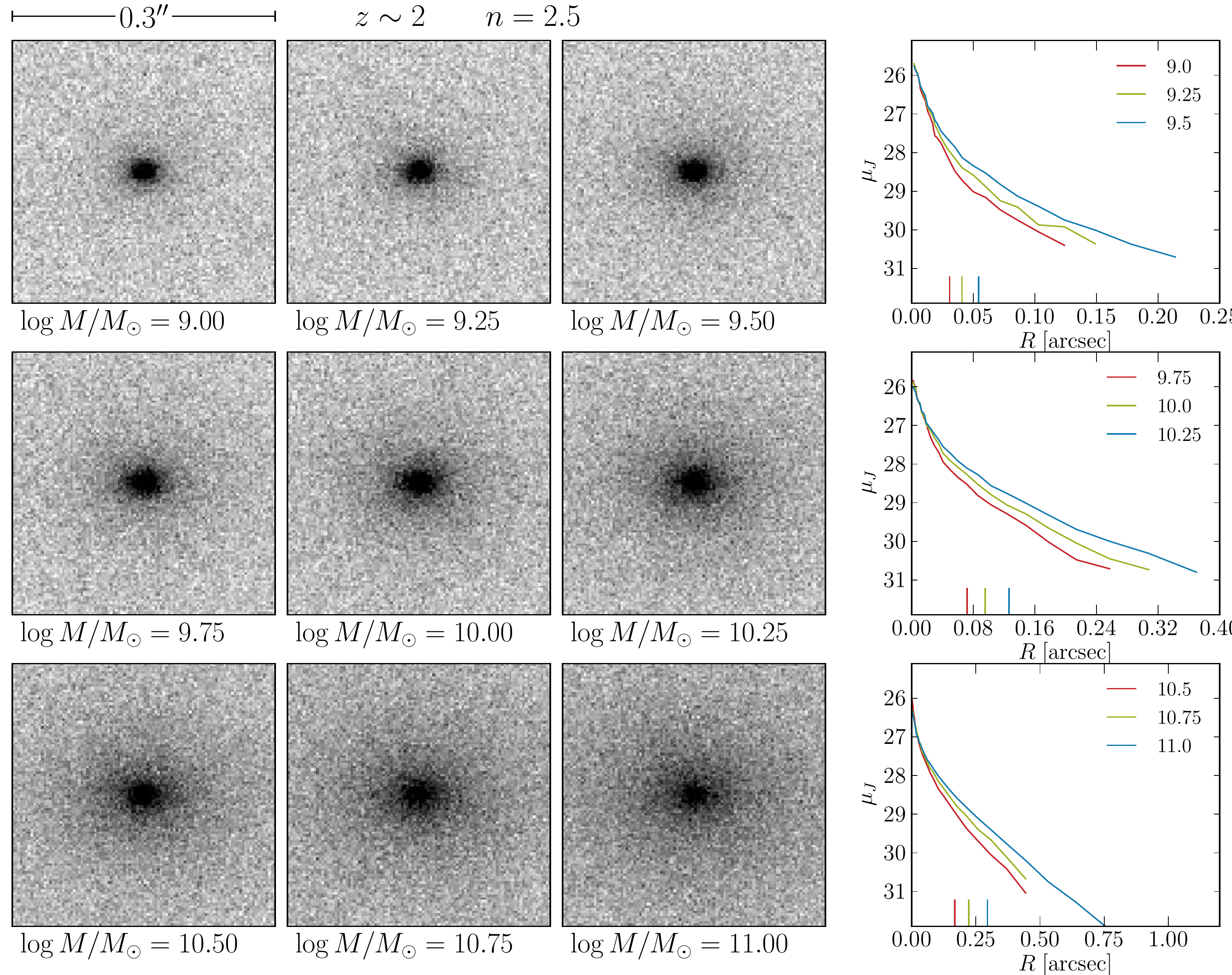}
  \caption{Central $0\farcs3 \times 0\farcs3$ region of  $J$-band images of a sample of simulated $z=2$ galaxies with Sersic index 
    $n=4.0$ and 2.5.
    The surface brightness profiles are shown in the right panels. The small vertical lines show the effective radii   of each galaxy.
  }
  \label{fig:imaz2_1}
\end{figure*}

\begin{figure*}
  \centering
  \includegraphics[width=14.7cm]{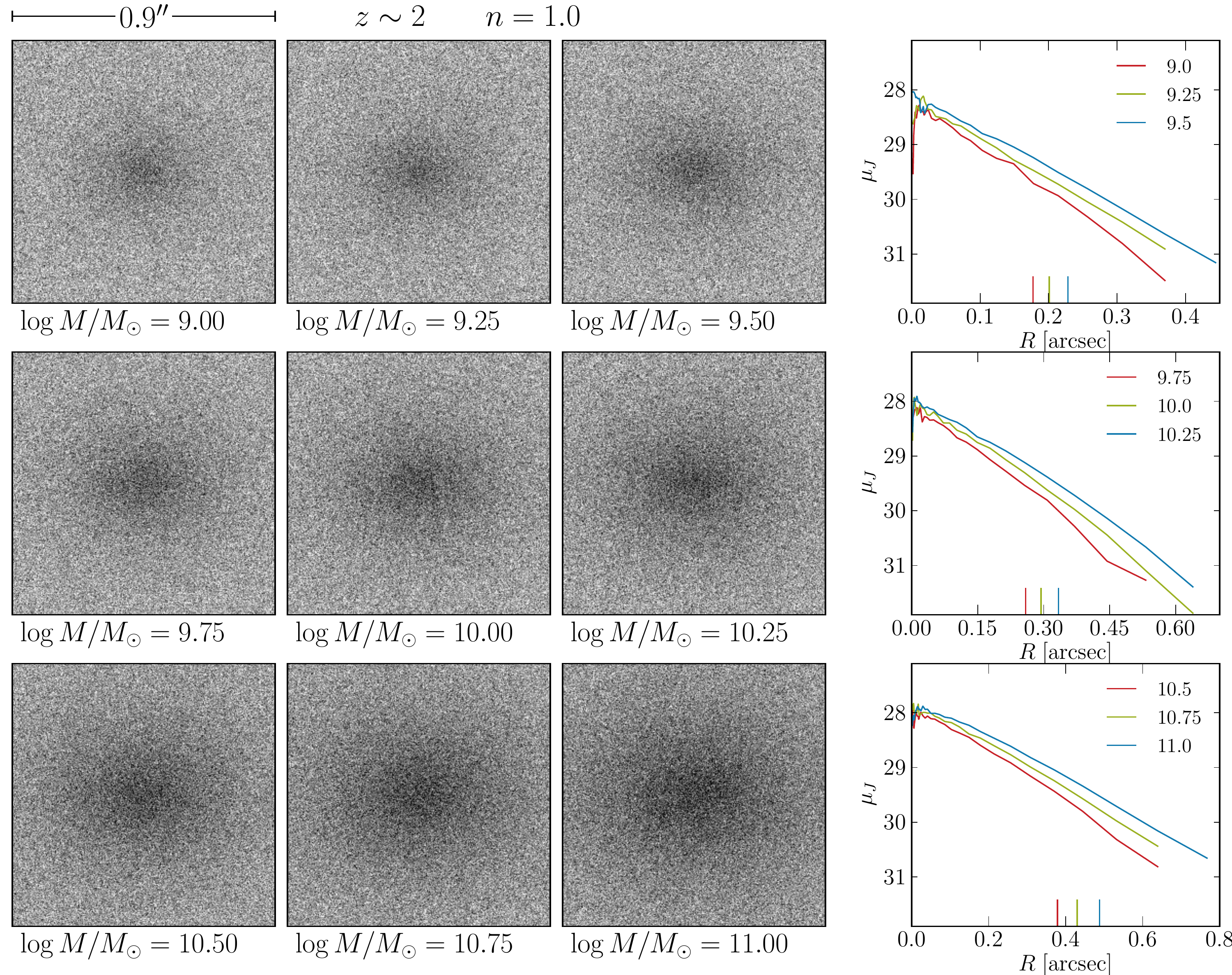}
  \caption{Central $0\farcs9 \times 0\farcs9$ region of $J$-band
    images of a sample of simulated $z=2$ galaxies. with Sersic index
    $n=1.0$.  The surface brightness profiles are shown in the right panels
The small vertical lines show the effective radii
    of each galaxy.  }
  \label{fig:imaz2_2}
\end{figure*}

Some examples of simulated galaxies of different mass and morphology and their azimuthally averaged radial surface brightness profiles
are shown in Fig. \ref{fig:imaz2_1} and \ref{fig:imaz2_2}.  The images for all simulated galaxies are available in the online
version of the paper in Fig. \ref{fig:imaHz2}, \ref{fig:imaHz3}, and \ref{fig:imaKz3}

\section{Results}

For each galaxy we measured the photometric (total magnitude) and structural parameters (effective radius $R_e$, Sersic index
$n$, ellipticity, and position angle) using GALFIT \citep{peng+2002}. 
The first guess for each parameter was randomly chosen from a distribution of values centred on the input value and with a dispersion of 30\%.
 We tested that even with first guess
parameters that deviate largely from the input values, the fitted values do not change  from those reported in this analysis.

A crucial ingredient for this analysis is the assumed PSF. In fact
since the size of the distant galaxies is very small (a fraction of arcsec), the effects of the PSF on the shape of the galaxies is
significant.  In this work, for the GALFIT measurements, we assumed the simulated image of a
bright star as PSF, which is obtained with AETC in the same configuration adopted for the galaxies.
This is clearly an ideal  situation and any additional source of instability (variable seeing, performance of the AO system, etc.) of the {\it real} 
observations will likely degrade the performances of the whole system.
We start by evaluating the accuracy of the measurements in this ideal
case; some effects due to variable PSF are
discussed in Sect. \ref{sec:varpsf}.

From the GALFIT analysis of each simulated  galaxy,  we measured the  $R_e$, $n,$ and total magnitude magnitude, and compared these values with the true 
 parameters. For each galaxy model we derive the mean and standard deviation of the difference of the parameters using the 50 independent runs. 
The results are given in  Tables \ref{tab:paramsz2} for the case at z = 2 and \ref{tab:paramsz3}, for z = 3 (available in the online version of the paper) 

\subsection{Galaxy structural parameters}\label{sec:params}

In Figure \ref{fig:paramsz2} and \ref{fig:paramsz3} we summarise the
results of the capability of determining the structural parameters
of high-redshift galaxies for different morphology (Sersic index) and
mass or size.
It turns out that, under the conditions described above,
with the combination of sensitivity and spatial resolution of
MICADO@EELT, it will be possible to characterise galaxies at z = 2 and
z = 3 with an accuracy better than $\sim$ 10\% and for objects with
masses as low as $10^9$ to $10^{10}$ $M_\sun$ (depending on
morphology and redshift; see details in Figure \ref{fig:paramsz2} and
\ref{fig:paramsz3}). The most critical objects are those at higher
redshift that are also more compact (high Sersic indices). Actually,
in both figures we only report the uncertainties of the galaxies with
size large enough to be measured.  The effective radius of early-type
galaxies ($n = 4$) at $ z=2$ is measured with an accuracy of $\sim
3\%$ (5\%) in the $H$ ($J$) band, for masses higher than $6\times10^9
M_\sun$. 
A similar accuracy is found for early-type galaxies at
redshift $z = 3$, provided that they are more massive than $10^{10}
M_\sun$. Objects for which these uncertainties apply are greater in size than $\sim 20$ mas, and their magnitudes and Sersic indices
are also very well determined.
 
For late-type galaxies ($n = 1$ and 2.5) at both redshifts, the
effective radius can be recovered with an accuracy better then 5 \%
over the whole explored mass range and actually better than 2\% for
the largest disks ($n=1$). The Sersic index and total magnitude of
late-type galaxies are also recovered with a good accuracy for all
masses at both $z = 2$ and 3 with a tendency to worsen at low masses
in the $z = 3$ case. At this high redshift, it appears that
measurements in the $K$ band will yield results that are less accurate (Fig. 5
right panel) owing to the relatively high contribution of the IR
background. Conversely, in the $H$ band, low mass, star-forming
galaxies with a shallow surface brightness profile ($n < 2.5$) have
effective radii of $ \sim 100$ mas and can be measured with an
accuracy better than $\sim 5$\%, however, their large size also
implies a fainter average surface brightness or a lower signal-to-noise ratio.  Nevertheless since these galaxies very often appear
clumpy their substructures will be relatively easy to detect (see
example in Sect. 3.5).  Since the galaxy population at high-$z$ is
expected to be dominated by disk/star-forming galaxies the assumed
imaging capabilities of MICADO will be adequate to study in detail the
bulk of galaxy population up to z $\sim$ 3 and beyond.  Besides the
statistical errors, we note some systematic deviations
between the average measured and  input value of galaxy parameters
(see Tables \ref{tab:paramsz2} and \ref{tab:paramsz3}). In most cases
these deviations are small compared with the estimated statistical (noise) error
but for the smallest objects. These systematic deviations may reflect
the numeric uncertainties in the PSF deconvolution when $R_e$ is of
the order of the size of the PSF core. The FWHM of MICADO PSF in the
$J$, $H$, and $K$ band is 2.5, 2.8, and 3.5 pixel (7.5, 8.4 and 10.5
mas), respectively. The $R_e$ of the galaxies are listed in Table
\ref{tab:scalerelz2} and \ref{tab:scalerelz3}.

\myfigure{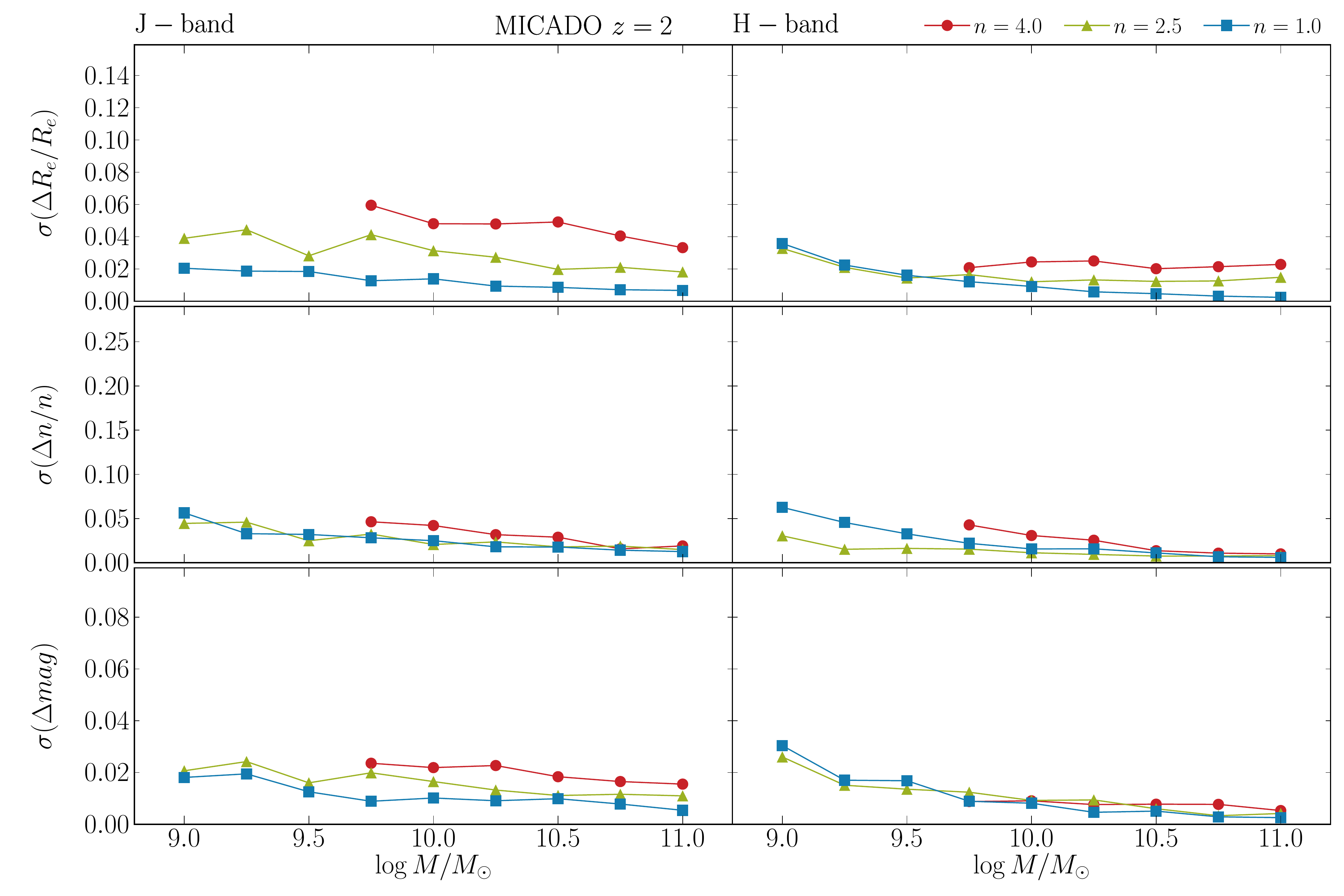}{
  Accuracy of structural parameters (effective radius, Sersic index, and total magnitude) measurements obtained from 
  MICADO observations of   $z=2$ galaxies with input Sersic indices equal to 1.0 ({\it blue squares}),   2.5 ({\it green triangles}), and 4.0 ({\it red circles}).
  Results obtained in the $J$ and $H$ bands are shown in the left and right panels, respectively.}{fig:paramsz2}

\myfigure{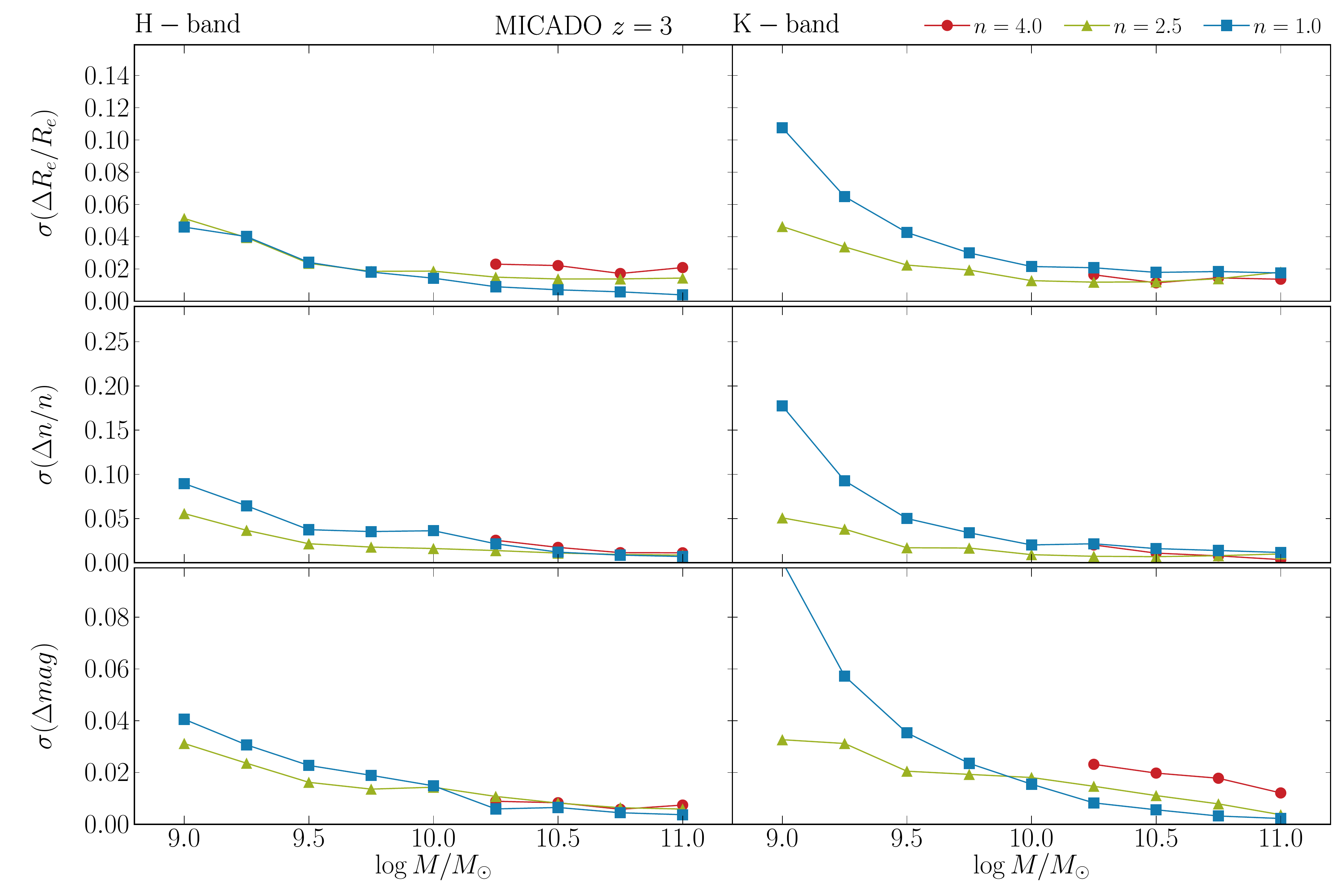}{
  As in Fig. \ref{fig:paramsz3} for $H$-and $K$-band simulations of $z=3$ galaxies.}{fig:paramsz3}

According to these  simulations,  MICADO should be able to provide accurate measurements of the structural and photometric
parameters of high-redshift galaxies that are impossible to achieve with current instrumentation. 
In fact a reliable measurement of the size of the smallest galaxies in our sample ($\lesssim 0\farcs1$) is
beyond the resolution limit of available space and ground-based cameras.
For instance, \cite{vand+2012} estimated the structural parameters of several high-z galaxies  measured from HST WFC3 IR  
observations of the CANDELS programme and found that an accuracy of $\sim$ 10\% is only obtained for objects 
brighter than $H=23$ mag. In order to properly measure the structure of fainter 
galaxies that represent the dominant fraction of the population, it is mandatory to use near-IR instrumentation with higher 
spatial resolution as that expected from ELTs.

 In conclusion  MICADO will provide very accurate measurements of the
structural parameters of high-redshift galaxies for all galaxies with  an angular size larger than 20 mas 
(corresponding to $\sim$ 0.15 kpc).  Even for galaxies with steep profiles ($n\sim4$) and very low mass, down to 
$10^9 M_\sun$, it will be possible to measure their effective radii and morphology (Sersic index) with uncertainty  of about 10--20 \%.

\subsection{Effects of PSF spatial variations}\label{sec:varpsf}

The simulations described above assume  ideal conditions.  Galaxies were simulated at the centre of  the MAORY FoV and GALFIT measurements were
carried out using  the simulated image of a star located  at same the position in the field as a PSF model.  Likely the PSF of real observations will be
affected by spatial and time variations. As a consequence, the PSF that will be used to analyse the galaxies profiles, will somehow be different from the PSF at the position of the galaxy.

To assess the uncertainty on the determination of structural parameters induced by PSF variations, one should explore a vast
parameter space, e.g. considering changes of Strehl ratio, variation of the elongation of the PSF core, and temporal and spatial variations. 
At the present stage of the design of the instruments it is not possible to properly
investigate the impact of  all these effects. This exercise is beyond the scope of this paper.  Nevertheless to shed some light on the
impact of PSF variation on this science case, we consider the specific circumstance in which the PSF model used in GALFIT to deconvolve the
galaxy radial surface brightness profile is not exactly the same as that used to generate the galaxy image with AETC. This test was carried out with three
different MAORY PSFs that were calculated for positions with an offset from the centre of the FOV of 5$\arcsec$, 10$\arcsec,$ and 20$\arcsec$ in both X and Y
directions.  We refer to these as PSF05, PSF10, and PSF20.

When using PSF05 and PSF10, we retrieved the correct $R_e$ with systematic
uncertainties of the order of 5-10\%, while using the PSF20 the uncertainties turn out of  20-30\%.  The error on the  magnitude
is always $\lesssim 0.1$ mag, while that  on the Sersic index depends on the size of the galaxies.  The systematic error
 in the measurements can be as high as 40\% for the smallest objects, but is of the order of 10\% for galaxies with $R_e >10$ pixels (40 mas).

We conclude that for science cases aiming at measuring structural parameters of relatively large galaxies ($R_e\sim40$ mas) with an
accuracy of $\sim 10\%$, PSF time and spatial variations would not be a critical issue. Particular care and/or dedicated observations would
instead be required to perform extremely accurate measurements of very compact galaxies.

\subsection{Comparison with NIRCAM at the JWST}

Before ELTs become operative, the most relevant future facility for the study of high-redshift galaxies is the NIRCam at the JWST. 
It is of interest thus to compare the expected performances of NIRCam at the JWST in the determinations of the structural
parameters of high-redshift galaxies with those expected for MICADO.
To this aim, we considered the galaxies at $z=2$ presented in the previous
sections.  We did not take into account galaxies with $n=4.0$ and $M<10^{10}M_\sun$ because their $R_e$  ($\sim$ 30 mas) is smaller than
1 pixel and it would therefore be beyond the limits of NIRCam spatial resolution\footnote{NIRCam pixel scale is 31.7 mas/pixels,
  i.e. $\sim 10$ times larger than that of MICADO.}.  All simulations were carried out with AETC using the NIRCam specifications given in
the NIRCam Exposure Time Calculator\footnote{version 1.6   \url{http://jwstetc.stsci.edu/etc/}}.

For these data we performed the same analysis as for MICADO simulations. The results are shown in Fig. \ref{fig:paramsz2JWST}.
For galaxies with $n=1$ the accuracy in the measurements in the bands J and H for $R_e$ and $n$ are $< 5\%$ over the whole range of 
considered parameters, and of $2-3\%$ for the total magnitude. These are star-forming galaxies that have $R_e$ in the range 200-500 mas and thus their global 
properties could be derived with similar accuracy by  JWST and E-ELT (see Figure  \ref{fig:paramsz2}  and \ref{fig:paramsz2JWST}   ). 
 Conversely, a significant  difference between MICADO and NIRCam accuracy is found for galaxies with $n=2.5$ and $n=4.0$ at $z =2$. 
 In the case of MICADO for all the simulated galaxies the accuracy of $R_e$ and $n$ remains better than $5\%,$ while  we found uncertainties in the range $10-25\%$ for NIRCam observations. 
 The accuracy is particularly poor for galaxies with $n=4$. This is because the $R_e$ of these galaxies is smaller than
$\sim2$ pixels in NIRCam images (see Table \ref{tab:scalerelz2}). 
 There is about a factor of 2 worse accuracy for galaxies with $n = 4$ compared to those with $n = 2.5$.
 Only for the most massive and largest galaxies ($M>10^{10.5}M_\sun$ and $R_e> 100$ mas ) the structural 
 parameters are recovered with an accuracy better than 5\%, but these represent  only a small fraction of the galaxy population. The estimated accuracy appears worse for $H$-band observations with respect to $J$-band observations.

\myfigure{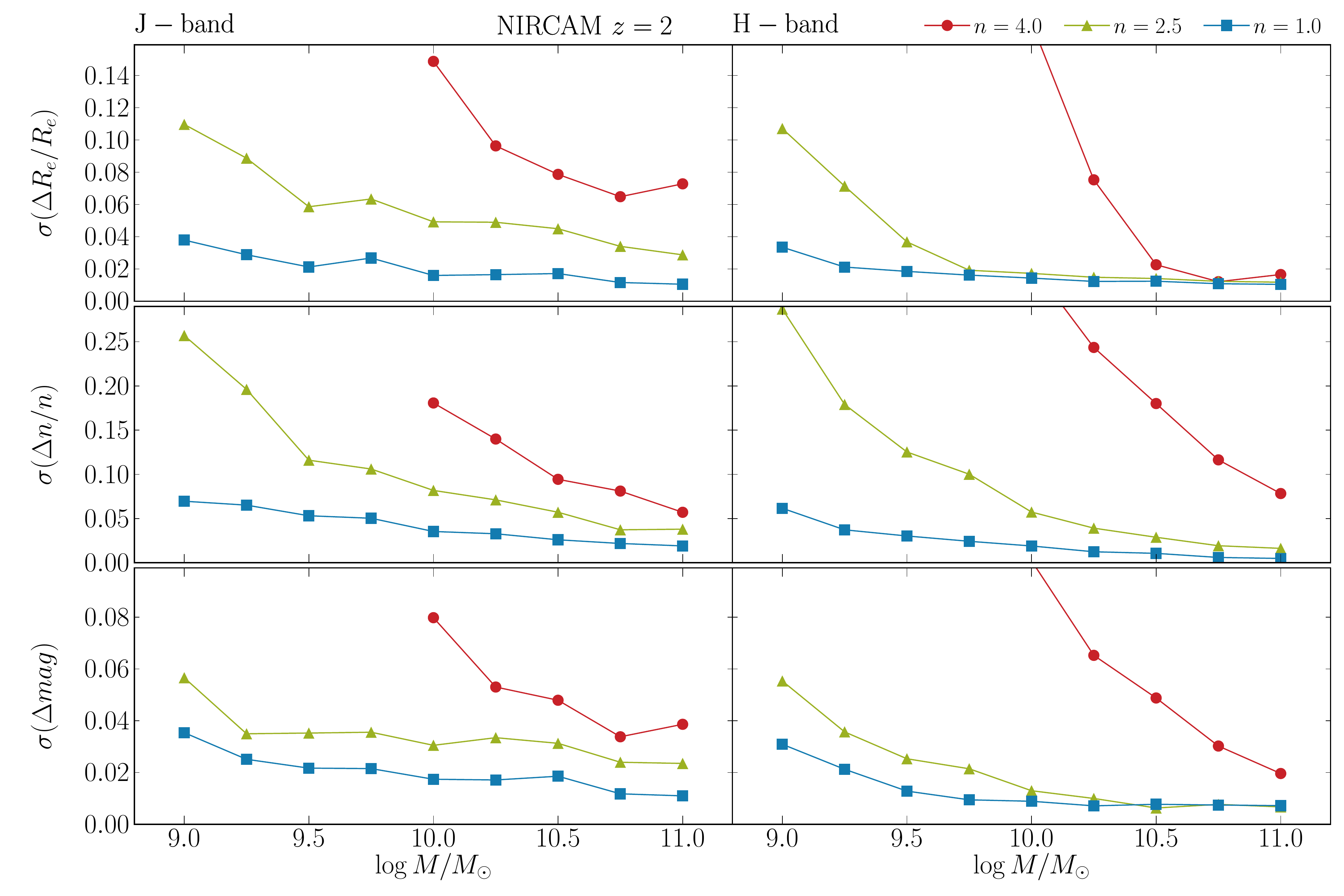}{
As in Fig \ref{fig:paramsz2} for simulations of NIRCcam at the JWST observations
of $J$-and $H$-band galaxies at $z=2$.}{fig:paramsz2JWST}

\subsection{Colour gradients}

In Sect. \ref{sec:params} we showed that with E-ELTs observations it
will be possible to characterise the properties of distant galaxies
with an accuracy of a few percent.  The comparison of high quality
images of these galaxies in different bands offers the opportunity to
investigate the trend of the colour of galaxies as a function of
distance from the centre. Colour gradients, and their dependence on
redshift, can be used to to test models of galaxy formation
\citep{sagl+2000,garg+2012}.  Decoding a colour gradient in terms of
an age or a metallicity gradient depends on the age and metallicity of
the stellar population. For relatively young stellar populations
($\lesssim 3$ Gyr), as appropriate for high-redshift galaxies, $\Delta
(U-V)/\Delta(\log(\mathrm{age})) \sim 1$ mag/dex at metallicities
around solar. The dependence of the U-V colour on metallicity is
lower, $\Delta(U-V)/\Delta\mathrm{[Fe/H]}\sim 0.3$ mag/dex, for
stellar populations of 1 to 3 Gyr of age, and metallicities between
$0.1\, Z_\sun$ and $2\, Z_\sun$.  Therefore, a U-V colour gradient of
0.1 magnitude may correspond to a variation of $\sim 25$\% in age or
of 0.3 dex in metallicity. The interpretation of colour gradients in
terms of the characteristics of the parent stellar population is
complicated by various effects, including the well-known,
age-metallicity degeneracy. It is however worth assessing the accuracy
with which it will be possible to measure the colour gradients of high-redshift galaxies, whose study will provide the means to investigate systematic variations of the stellar populations.

In these simulations we used the same input structural parameters
($R_e$ and $n$) in the two photometric bands for all galaxies.
Therefore, our simulated galaxies have colour gradients that are equal to zero
by construction.  The errors in the output structural parameters introduce a spurious colour gradient, which can be considered an
error on the colour gradient measured from the simulated images.
Since the accuracy of the derived colour profile does not strongly
depend on the actual value of the colour gradient, but rather on the
total magnitude of the galaxy and its size, the uncertainties derived
from the analysis of galaxies with zero colour gradient are also valid
for galaxies with non-zero colour gradient.

We computed the rest-frame $U-V$ colour
profile as the difference between the best-fit Sersic model in $J$-
and $H$-band for galaxies at $z=2$ and $H$- and $K$-band at $z=3$ for each simulated galaxy. The
gradient was then calculated by fitting a linear relation to the
colour profile ( in the $\mu_U-\mu_V$ versus $\log R$ plane) in the region
$0.1<R/R_e<1.0$.  We evaluated the mean and standard deviation values
of the distributions of the 50 gradients that we computed for the 50
simulations of each template galaxy in each band.  The results are
reported in Tables \ref{tab:gradsz2} and \ref{tab:gradsz3}.  As in the case of the structural
parameters, we evaluated the overall uncertainty in the colour
gradient as the sum in quadrature of the systematic error and of the
random error (see Fig. \ref{fig:gradsz2z3}).  The recovered colour
gradients for galaxies with $n=1.0$ and 2.5 are all compatible, within
the uncertainties, with the input value (zero) for both $z=2$ and
$z=3$ galaxies.

At $z=2$, the uncertainties in the colour gradient measurements are of
the order of $\lesssim0.05$ mag/dex for galaxies with
$M\gtrsim10^{10}M_\sun$ and $\sim 0.10$ mag/dex for the smallest
galaxies with $M=10^{9}M_\sun$ (upper panel in
Fig. \ref{fig:gradsz2z3}).  The results are essentially the same for
galaxies at $z=3$ with $n=4.0$ and 2.5. The uncertainties are larger
for late-type galaxies ($n=1.0$).  They are of the order of $\sim
0.10$ mag/dex for galaxies with $M\gtrsim10^{10}M_\sun$ and up to
$\sim 0.35$ mag/dex for $M=10^9M_\sun$ galaxies. This is mainly because of the lower accuracy of the structural parameters derived from from
$K$-band images (see Fig. \ref{fig:paramsz3}).

In summary we found that MICADO images of high-z galaxies can be used
to measure their colour gradients in the inner galactic regions
($R<R_e$) with an accuracy of 0.1 mag/dex in all galaxies with
$M\gtrsim 10^9 M_\sun$ at $z=2$ and with $M\gtrsim 10^{10} M_\sun$ at
$z=3$. A similar accuracy is also expected for early-type galaxies
($n\ge2.5$) at $z=3$ and $M\gtrsim 10^{9} M_\sun$. In the case of
late-type dwarf galaxies at $z=3$ we instead expect an accuracy of
about of 0.3 mag/dex.

\myfigure{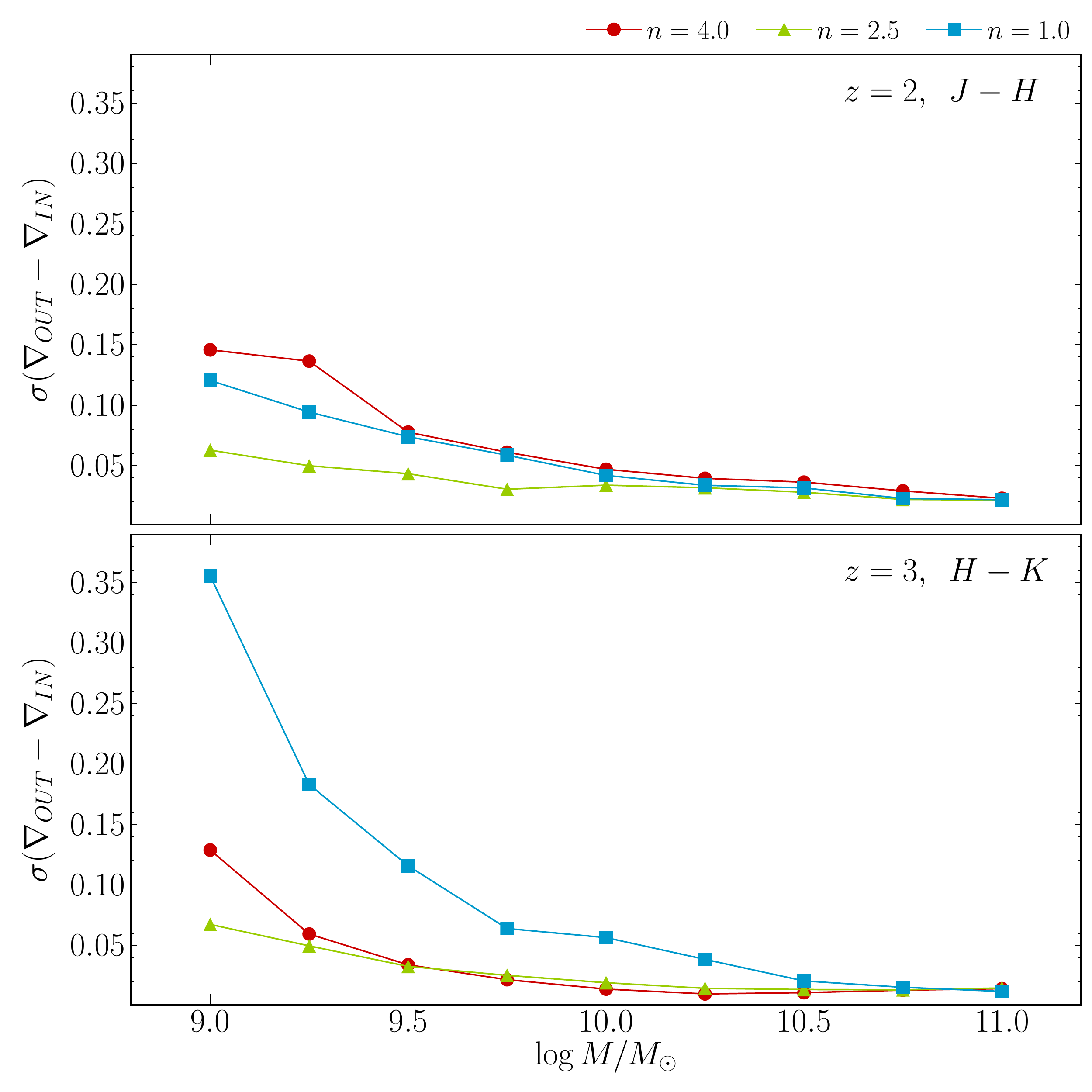}{
Uncertainty in the colour gradient for galaxies of different stellar mass and Sersic index, as obtained from simulations of 
MICADO observations of high-redshift galaxies. The $J-H$ and $H-K$ colours correspond to rest-frame $U-V$ at
redshift  $z=2$  ({\it top panel}) and  $z=3$  ({\it bottom panel}).
}{fig:gradsz2z3}

\subsection{Simulated  high-$z$ galaxies using templates }

While for early-type passive galaxies the Sersic law provides an
accurate description of the surface brightness profile, late-type star-forming galaxies have a much more complex morphology. Besides the
exponential stellar disk, in late-type galaxies there are often also a
number of conspicuous substructures, such as spiral arms, clumpy star-forming regions, and dust lanes. The study of these features offers
further important clues to understand the processes of galaxy
formation and evolution.  In this section we complement the
description of the capabilities of ELTs and JWST to characterise the
morphology and structural parameters of galaxies with simulations of
observations of galaxies with a more complex morphology.  Simulations
of observations of high-$z$ galaxies using HARMONI, the future
integral-field spectrograph for the E-ELT, was presented by
\cite{kend+2016}.

These kinds of simulations are of particular interest when studying
galaxies at $z\simeq2-3$ because these redshifts correspond to the
epoch of maximum cosmic star formation rate \citep{mada+2014}, when
the fraction of star-forming late-type galaxies and merging galaxies
was much higher than at the present epoch.

\begin{figure*}
    \centering
    \includegraphics[width=12.2cm]{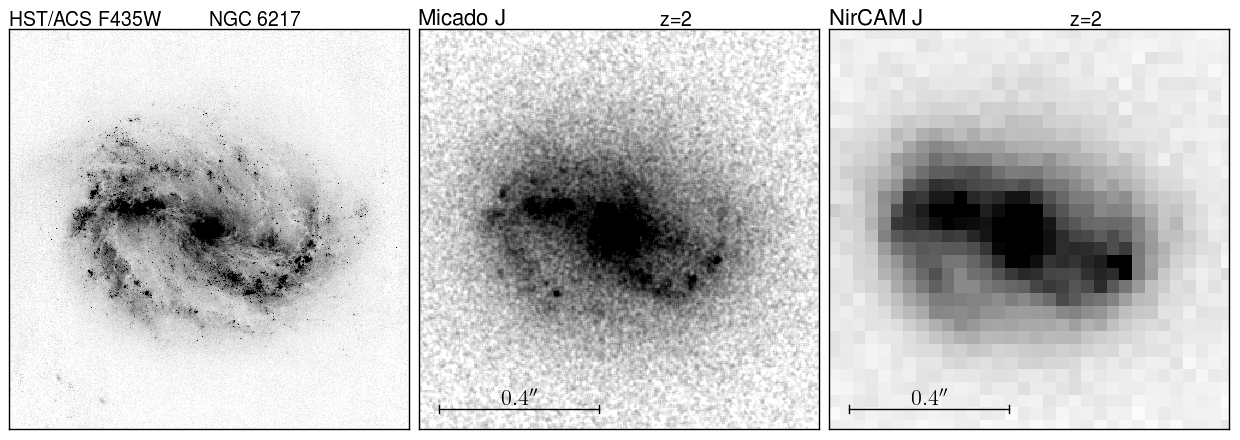}\\
    \includegraphics[width=12.2cm]{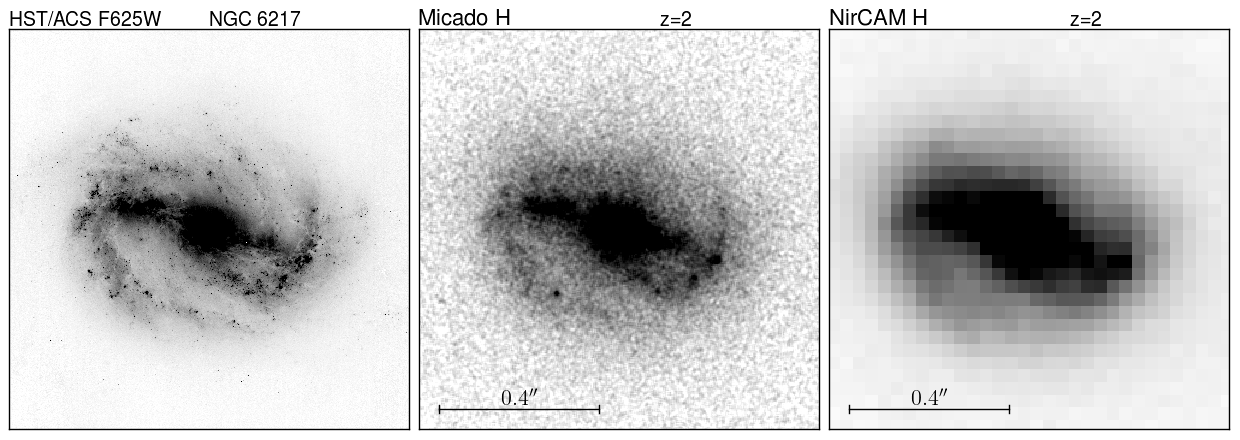}\\\vspace{1em}
    \includegraphics[width=12.2cm]{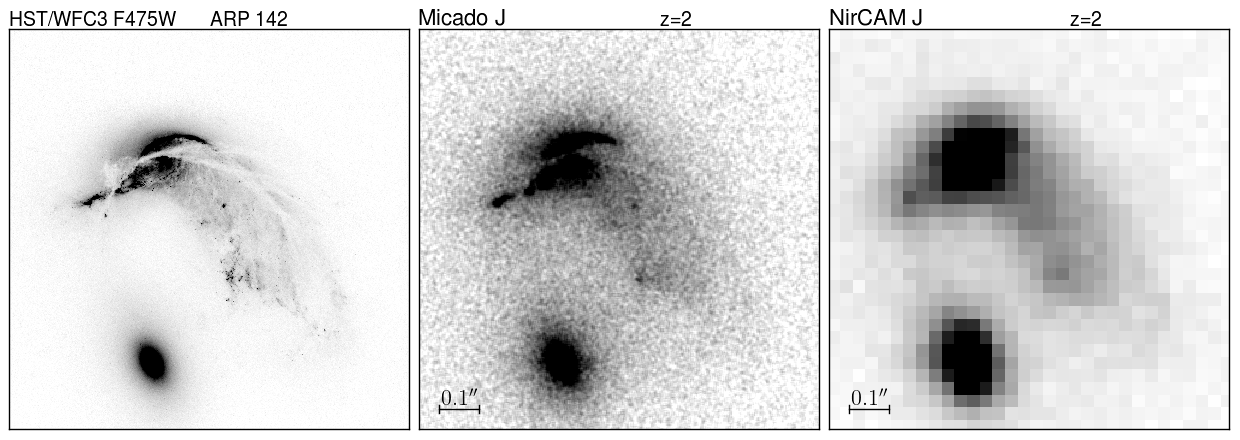}\\
    \includegraphics[width=12.2cm]{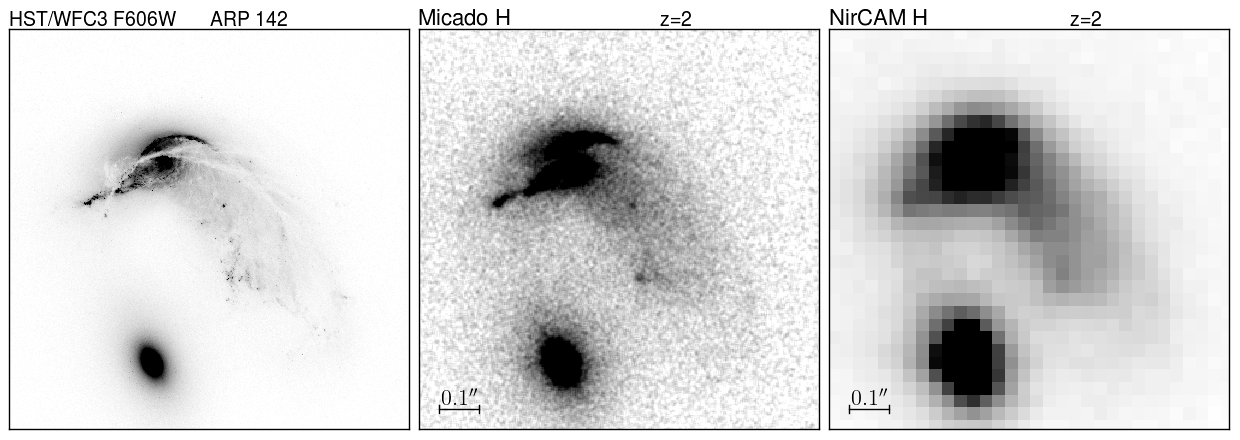}\\\vspace{1em}
    \includegraphics[width=12.2cm]{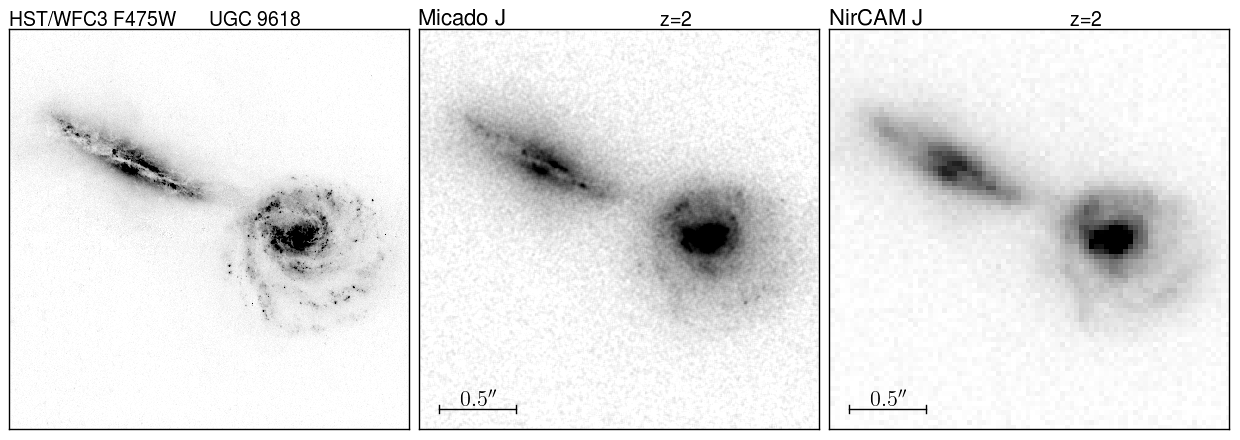}\\
    \caption{Simulations of MICADO ({\it central panels}) and NIRCam ({\it right panels}) $J$- and $H$-band observations of galaxies at $z\sim2$  obtained using HST observations of nearby galaxies as templates ({\it left panels})}.
    \label{fig:templatesz2}
  \end{figure*}

\begin{figure*}
    \centering
    \includegraphics[width=12.2cm]{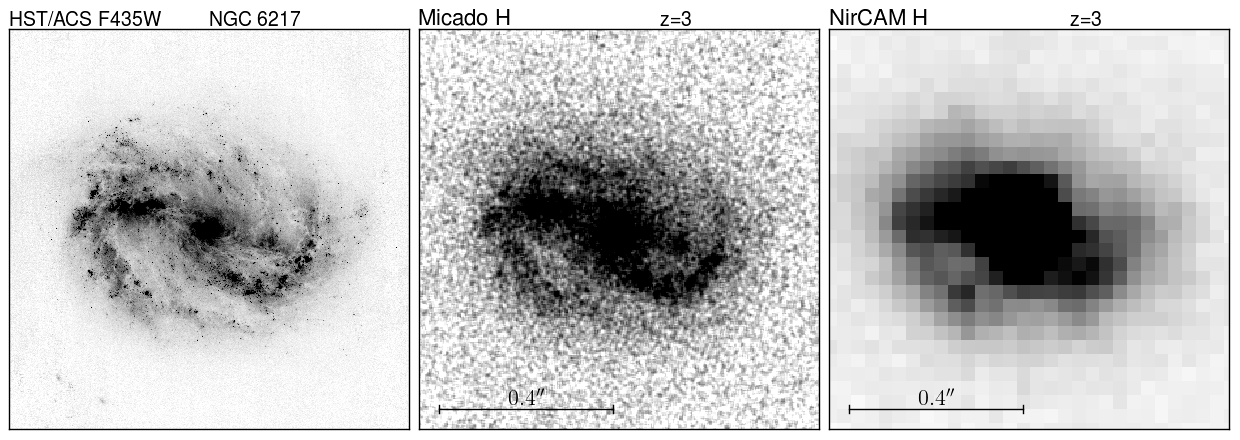}\\
    \includegraphics[width=12.2cm]{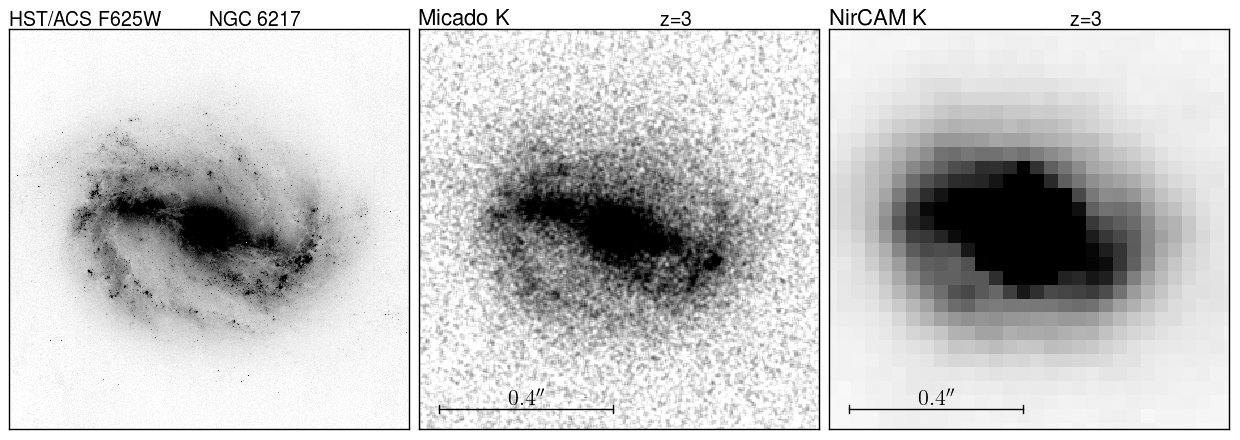}\\\vspace{1em}
    \includegraphics[width=12.2cm]{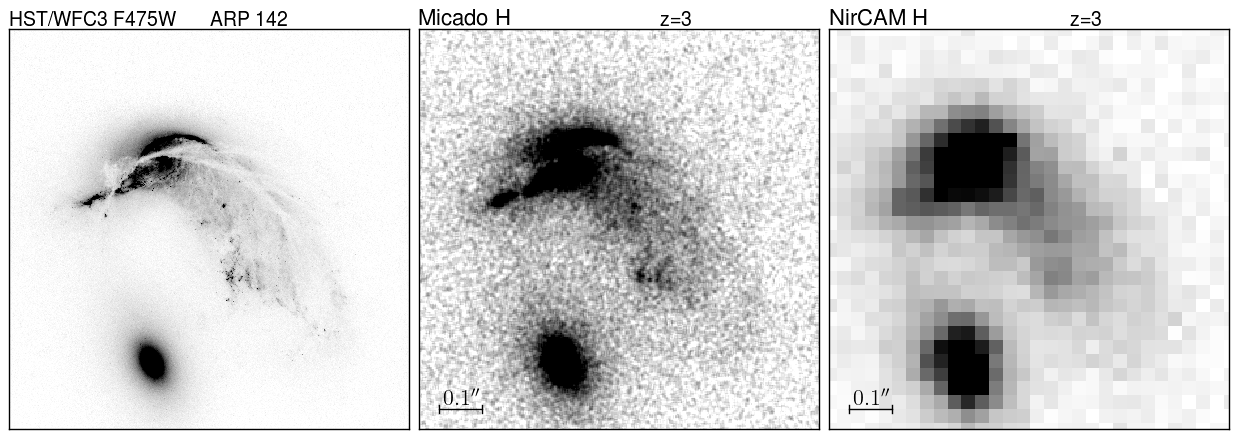}\\
    \includegraphics[width=12.2cm]{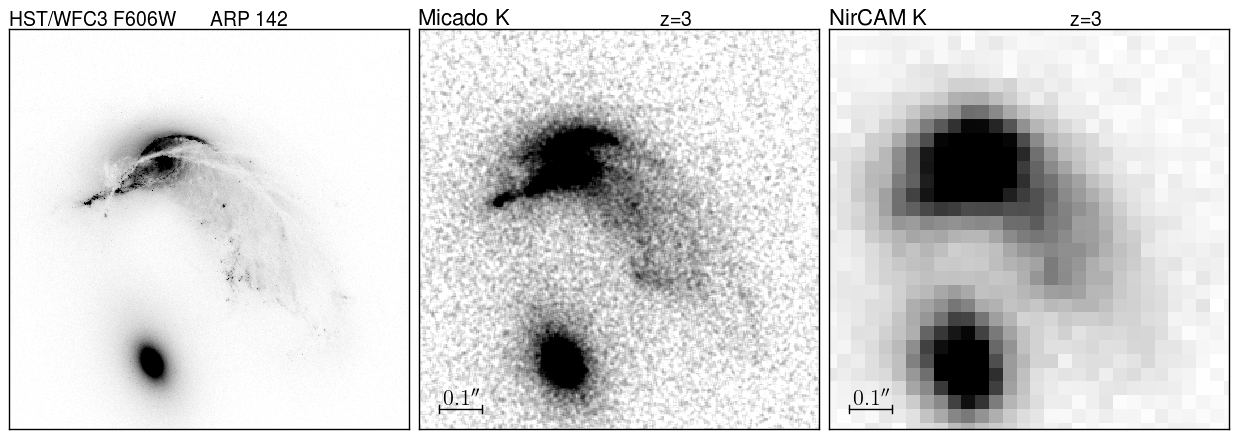}\\\vspace{1em}
    \caption{Simulations of MICADO ({\it central panels}) and NIRCam ({\it right panels}) $H$- and $K$-band observations of galaxies at $z\sim3$  obtained using HST observations of nearby galaxies as templates ({\it left panels})}.
    \label{fig:templatesz3}
  \end{figure*}

We performed simulated observations of high-$z$ galaxies with template images obtained from high-resolution observations of nearby
galaxies.  To this aim, we looked for Advance Camera for Survey (ACS)
and/or WFC3 observations of galaxies with different morphologies in
the HST archive.  We selected observations in photometric bands that
roughly correspond to $J$ and $H$ ($H$ and $K$) when redshifted at
$z\sim2$ ($\sim3$) for three different objects: (i) images in the
filters F435W and F606W of NGC~6217 ($z=0.004$), a star-forming barred
spiral galaxy; (ii) image in filter F435W of UGC~9618 ($z=0.033$), a
pair of gas-rich galaxies (the first is a face-on spiral and the
second is a edge-on galaxy with a dense dust lane) in the early
stage of interaction; and (iii) images in filters F475W and F625W of
ARP~142 ($z=0.023$), which is a violent merger of an elliptical galaxy with a
disrupting gas-rich galaxy.  We removed
all foreground stars, background galaxies, cosmic rays, and artefacts from the original HST images.
We then used AETC to produce MICADO and JWST simulated observations
(using same prescriptions described in Sect.  2) with a total
integration time of three hours. The total flux and angular size of
the templates were rescaled to match the magnitude and size of
galaxies at redshift 2 and 3, using the values in
Fig. \ref{fig:scalerel} as a reference.  These simulated observations
are shown in Fig. \ref{fig:templatesz2} and \ref{fig:templatesz3}.
Each set of three figures shows the original template the MICADO image
and the NIRCam image. We applied a a box car filter with size of 3
pixels to the MICADO images to increase the signal-to-noise ratio to
enhance the low surface brightness features.

Since the spatial resolution of MICADO is a factor $\sim6$ higher than
that of NIRCam, these small substructures are much better defined in
MICADO images than in NIRCam images. For example, JWST could barely
resolve the bar and spiral arms of a late-type galaxy with $R_e$
$\sim$ $0\farcs3$, while a 40-m class ground-based ELT would also be able to
resolve the brightest star-forming regions; see e.g. the
simulations obtained using NGC~6217 and UGC~9618 as a template Figures
\ref{fig:templatesz2} and \ref{fig:templatesz3}.  ELTs would also be
able to detect dense dust lanes in galaxies up to redshift 2 or 3.
The third example was obtained using the violent merger ARP~142 and
shows that, in spite of the significant lower background for JWST
observations in $H$ and $K$ bands, the combination of higher
resolution and larger sensitivity (39 m aperture compared with 6.5 m) of
E-ELT allow MICADO to detect the low surface brightness regions (
e.g. the tail of the disrupting late-type galaxy in ARP~142) slightly
better than NIRCam and also to distinguish the fine ($\sim$ 50 mas)
substructures (see Figures \ref{fig:templatesz2} and
\ref{fig:templatesz3}).


\section{Summary and conclusions}

In this paper we presented a detailed analysis of the expected performances of next-generation 
ELTs in the characterisation of the properties of high-$z$ galaxies. We evaluated the accuracy in the measurements
of the structural and photometric parameters that will be obtained from future observations
using an extensive set of simulated observations. As a reference, we used the specifications of the MICADO camera
to be mounted at the E-ELT and the NIRCAM at the JWST. The extraordinary large diameter of the E-ELT
will provide imaging of high-redshift galaxies with unprecedented spatial resolution.

Our results confirm that MICADO will provide extremely accurate measurements of the structural parameters of high-redshift galaxies;
systematic uncertainties in the measurements of structural parameters are expected to be negligible for galaxies with effective radii larger
than $\sim20$ mas. The overall uncertainties in the measurements of the structural parameters of late-type galaxies result to be smaller than
$5\%$ and the uncertainties in the total magnitude of a few hundredths of magnitude, even for the smallest simulated galaxies ($M=10^9 M_\sun$, $H\sim$24.5 mag).
For early-type galaxies that are more compact and have a steeper profile ($n=4$) than late-type galaxies,
we found that MICADO would provide measurements of the effective radius and the Sersic index with uncertainties
of just 10--20\% even for the smallest and faintest galaxies considered in our study ($M=10^9 M_\sun$, $H\sim$26.5 mag).
These results confirm that MICADO observations will represent a real breakthrough as they will facilitate results far 
beyond the capabilities of present-day instrumentation. Today, this accuracy is obtained with the HST for
galaxies $\sim 3$ mag brighter and at least 10 times more massive and/or bigger \citep[see e.g.][]{vand+2014}.

We also estimate that, from ELTs observations, it will be possible to obtain reliable measures of
colour gradients as small as 0.3 mag/dex in galaxies at $z=3$. Therefore it will be possible 
to study the population gradients in high-redshift galaxies. This will represent a fundamental step
to understand the assembly history and  physical processes regulating the formation, the growth 
and, evolution of the galaxies. This study is nowadays possible only for nearby galaxies; 
the advent of the ELTs will make it possible to extend these studies at $z=2$ and 3, 
probing galaxies at the epoch in which the star formation rate density peaked \citep[$z\simeq2$; e.g.][]{mada+2014} and beyond.

We also provided a quantitative estimate of the performances expected for the JWST.
We found that NIRCam will provide excellent measurements for late-type galaxies down to 
$M\sim10^{10} M_\sun$ at $z$=2. The size of low-mass early-type galaxies ($M\lesssim10^{10}M_\sun$)
at $z=3$ would be of the order of (or even smaller than) the spatial resolution of the  JWST and therefore
it would be extremely difficult to derive reliable measurements of their spatial and photometric parameters.
The first opportunity to explore the details of the structure of distant compact galaxies, or the very inner region
of galaxies at lower redshift, will be offered by the advent in the next decade of 
the imagers at ground-based extremely large telescopes and, in particular, of MICADO at the E-ELT.

We finally performed simulations of high-$z$ galaxies using high-resolution HST images of galaxies at low redshift.
These simulations manifest  how the combined capabilities of excellent angular resolution and sensitivity will allow us to investigate 
in great detail the small substructures (spiral arms, clumpy star-forming regions, dust lanes, etc.) of distant galaxies that could not be investigated by any other 
present or actually planned ground- or space-based facilities.

\begin{acknowledgements}
We acknowledge  support of INAF and MIUR through the grant {\it Progetto premiale T-REX}.
\end{acknowledgements}

\bibliographystyle{aa}
\bibliography{gullieuszik}

\Online
\begin{appendix}
\section{Supplementary tables and figures}

\begin{table}[!ht]
    \caption{Effective radius, $J$- and $H$-band magnitudes adopted for the template galaxies at $z=2$.}
    \label{tab:scalerelz2}
    \centering            
    \begin{tabular}{ccccc}
      \hline\hline         
      $\log M/M_\odot$&
      $R_e$ [kpc]&
      $R_e$ [$\arcsec$]&
      $J$&$H$\\   
      \hline
\multicolumn{5}{c}{$n=1.0$} \\
 9.00 &   1.5005 &   0.1772 & 25.000 & 25.000 \\ 
 9.25 &   1.7030 &   0.2011 & 24.726 & 24.538 \\ 
 9.50 &   1.9330 &   0.2283 & 24.450 & 24.075 \\ 
 9.75 &   2.1939 &   0.2591 & 24.176 & 23.613 \\ 
10.00 &   2.4902 &   0.2941 & 23.900 & 23.150 \\ 
10.25 &   2.8264 &   0.3338 & 23.626 & 22.688 \\ 
10.50 &   3.2079 &   0.3788 & 23.350 & 22.225 \\ 
10.75 &   3.6410 &   0.4300 & 23.076 & 21.763 \\ 
11.00 &   4.1326 &   0.4880 & 22.800 & 21.300 \\ 
\multicolumn{5}{c}{$n=2.5$} \\
 9.00 &   0.2615 &   0.0309 & 26.125 & 25.500 \\ 
 9.25 &   0.3467 &   0.0409 & 25.788 & 25.038 \\ 
 9.50 &   0.4597 &   0.0543 & 25.450 & 24.575 \\ 
 9.75 &   0.6095 &   0.0720 & 25.113 & 24.113 \\ 
10.00 &   0.8082 &   0.0954 & 24.775 & 23.650 \\ 
10.25 &   1.0715 &   0.1265 & 24.438 & 23.188 \\ 
10.50 &   1.4207 &   0.1678 & 24.100 & 22.725 \\ 
10.75 &   1.8837 &   0.2224 & 23.763 & 22.263 \\ 
11.00 &   2.4975 &   0.2949 & 23.425 & 21.800 \\ 
\multicolumn{5}{c}{$n=4.0$} \\
 9.00 &   0.0456 &   0.0054 & 27.250 & 26.000 \\ 
 9.25 &   0.0706 &   0.0083 & 26.850 & 25.538 \\ 
 9.50 &   0.1093 &   0.0129 & 26.450 & 25.075 \\ 
 9.75 &   0.1693 &   0.0200 & 26.050 & 24.613 \\ 
10.00 &   0.2623 &   0.0310 & 25.650 & 24.150 \\ 
10.25 &   0.4062 &   0.0480 & 25.250 & 23.688 \\ 
10.50 &   0.6292 &   0.0743 & 24.850 & 23.225 \\ 
10.75 &   0.9745 &   0.1151 & 24.450 & 22.763 \\ 
11.00 &   1.5093 &   0.1782 & 24.050 & 22.300 \\
\hline
    \end{tabular}
  \end{table}

\begin{table}[!ht]
    \caption{Effective radius, $H$- and $K$-band magnitudes adopted for the template galaxies at $z=3$.}
    \label{tab:scalerelz3}
    \centering            
    \begin{tabular}{ccccc}
      \hline\hline        
      $\log M/M_\odot$&
      $R_e$ [kpc]&
      $R_e$ [$\arcsec$]&
      $H$&$K$\\    
      \hline
\multicolumn{5}{c}{$n=1.0$} \\
 9.00 &   1.2604 &   0.1634 & 25.500 & 25.800 \\ 
 9.25 &   1.4306 &   0.1855 & 25.038 & 25.150 \\ 
 9.50 &   1.6237 &   0.2106 & 24.575 & 24.500 \\ 
 9.75 &   1.8430 &   0.2390 & 24.113 & 23.850 \\ 
10.00 &   2.0918 &   0.2712 & 23.650 & 23.200 \\ 
10.25 &   2.3742 &   0.3079 & 23.188 & 22.550 \\ 
10.50 &   2.6947 &   0.3494 & 22.725 & 21.900 \\ 
10.75 &   3.0585 &   0.3966 & 22.263 & 21.250 \\ 
11.00 &   3.4715 &   0.4502 & 21.800 & 20.600 \\ 
\multicolumn{5}{c}{$n=2.5$} \\
 9.00 &   0.2082 &   0.0270 & 26.000 & 25.675 \\ 
 9.25 &   0.2752 &   0.0357 & 25.538 & 25.088 \\ 
 9.50 &   0.3639 &   0.0472 & 25.075 & 24.500 \\ 
 9.75 &   0.4811 &   0.0624 & 24.613 & 23.913 \\ 
10.00 &   0.6360 &   0.0825 & 24.150 & 23.325 \\ 
10.25 &   0.8409 &   0.1090 & 23.688 & 22.738 \\ 
10.50 &   1.1117 &   0.1442 & 23.225 & 22.150 \\ 
10.75 &   1.4697 &   0.1906 & 22.763 & 21.563 \\ 
11.00 &   1.9430 &   0.2520 & 22.300 & 20.975 \\ 
\multicolumn{5}{c}{$n=4.0$} \\
 9.00 &   0.0344 &   0.0045 & 26.500 & 25.550 \\ 
 9.25 &   0.0530 &   0.0069 & 26.038 & 25.026 \\ 
 9.50 &   0.0816 &   0.0106 & 25.575 & 24.500 \\ 
 9.75 &   0.1256 &   0.0163 & 25.113 & 23.976 \\ 
10.00 &   0.1934 &   0.0251 & 24.650 & 23.450 \\ 
10.25 &   0.2978 &   0.0386 & 24.188 & 22.926 \\ 
10.50 &   0.4586 &   0.0595 & 23.725 & 22.400 \\ 
10.75 &   0.7062 &   0.0916 & 23.263 & 21.876 \\ 
11.00 &   1.0875 &   0.1410 & 22.800 & 21.350 \\ 
\hline
    \end{tabular}
  \end{table}

\begin{figure}
  \centering
  \includegraphics[width=\hsize]{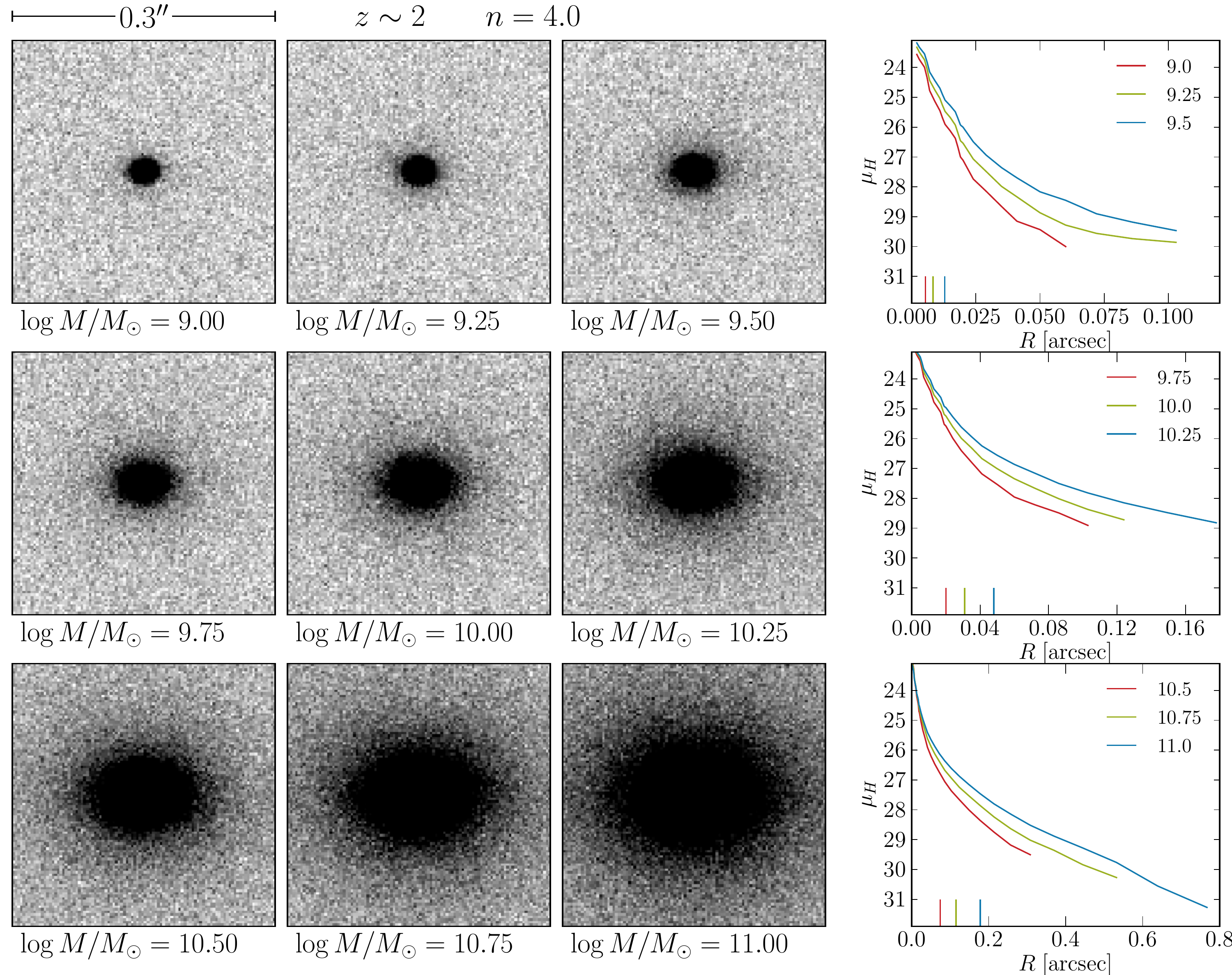}\\[1em]
  \includegraphics[width=\hsize]{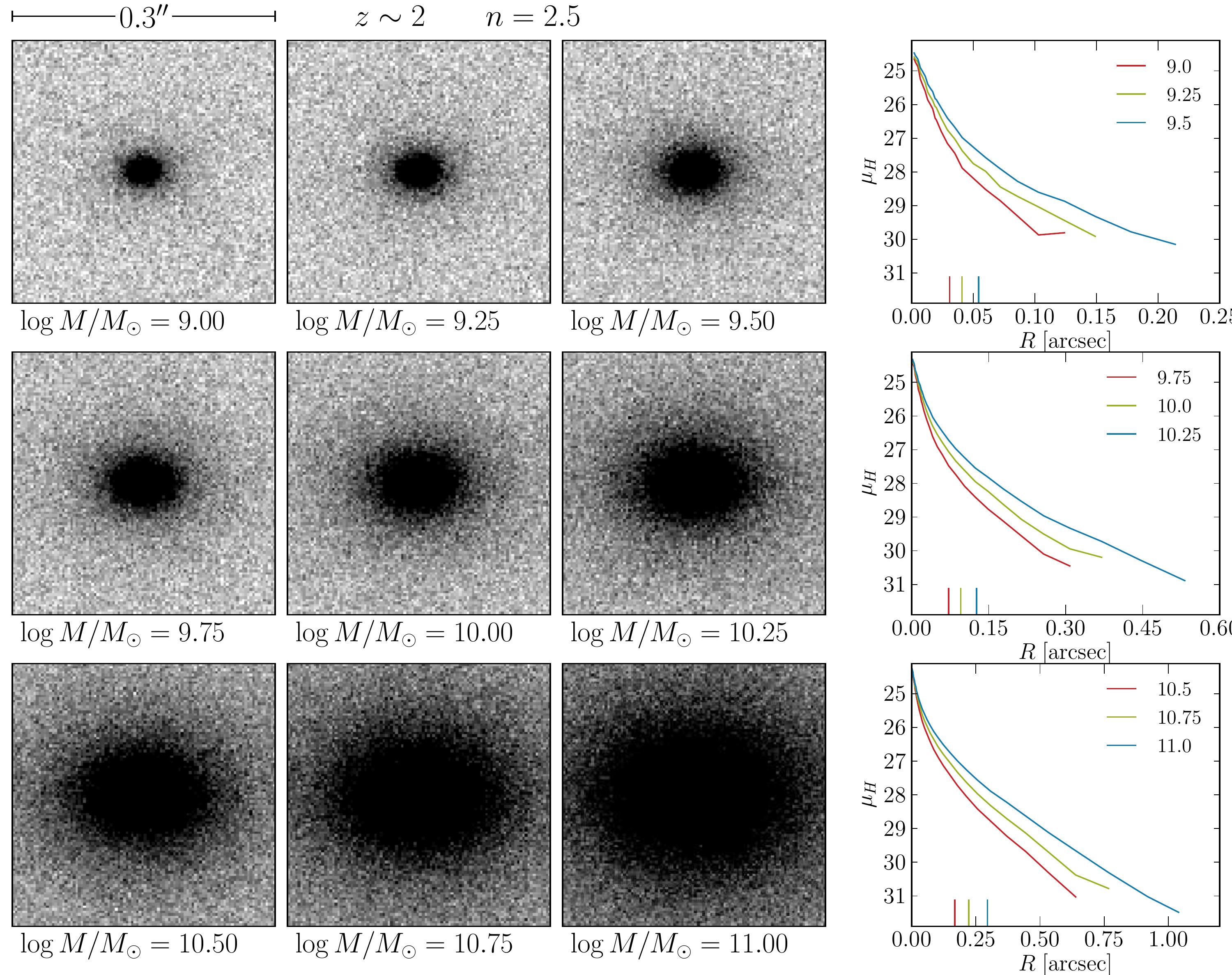}\\[1em]
  \includegraphics[width=\hsize]{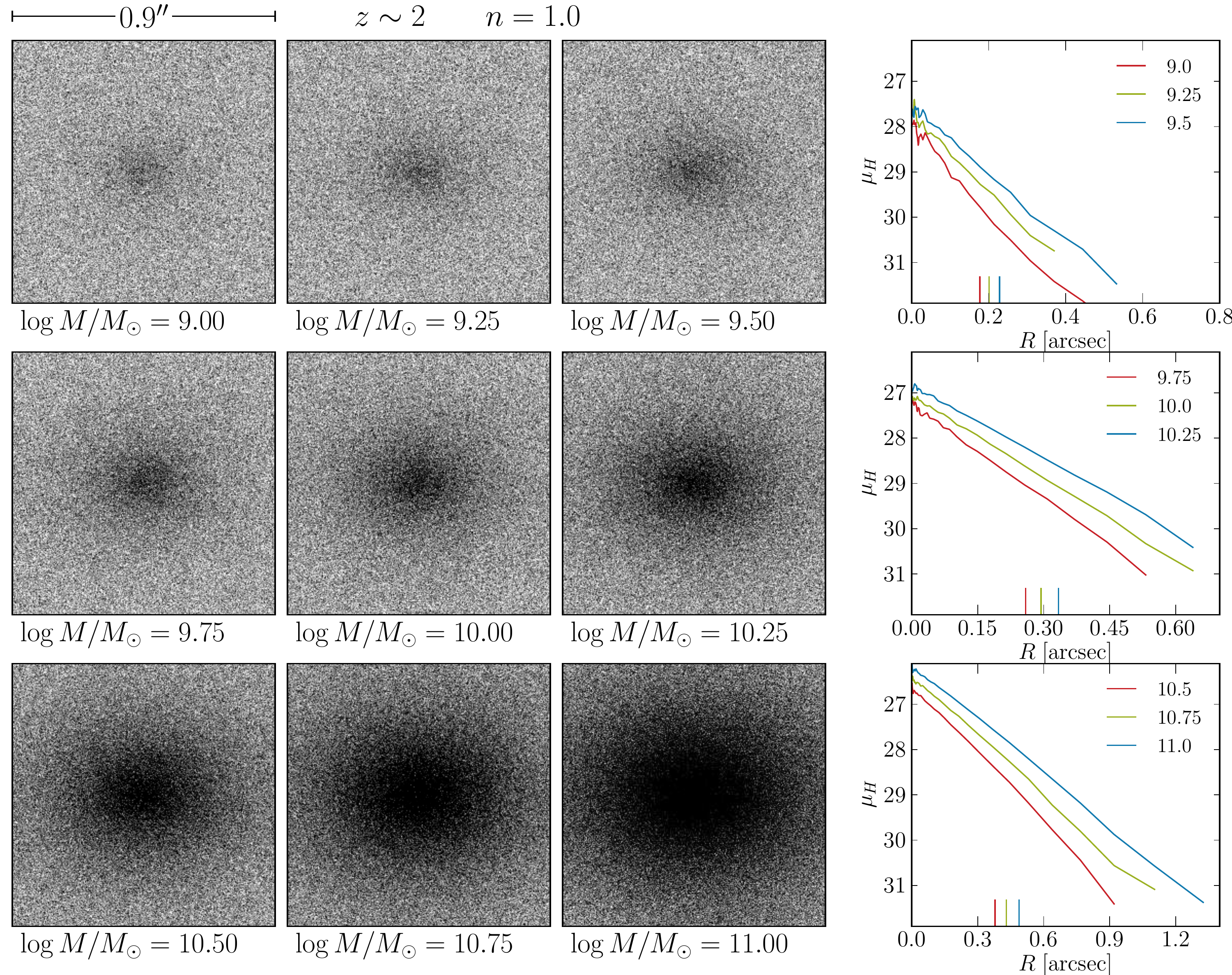}
  \caption{As in \ref{fig:imaz2_1} for $H$-band images of  $z=2$ galaxies.}
  \label{fig:imaHz2}
\end{figure}
\begin{figure}
  \centering
  \includegraphics[width=\hsize]{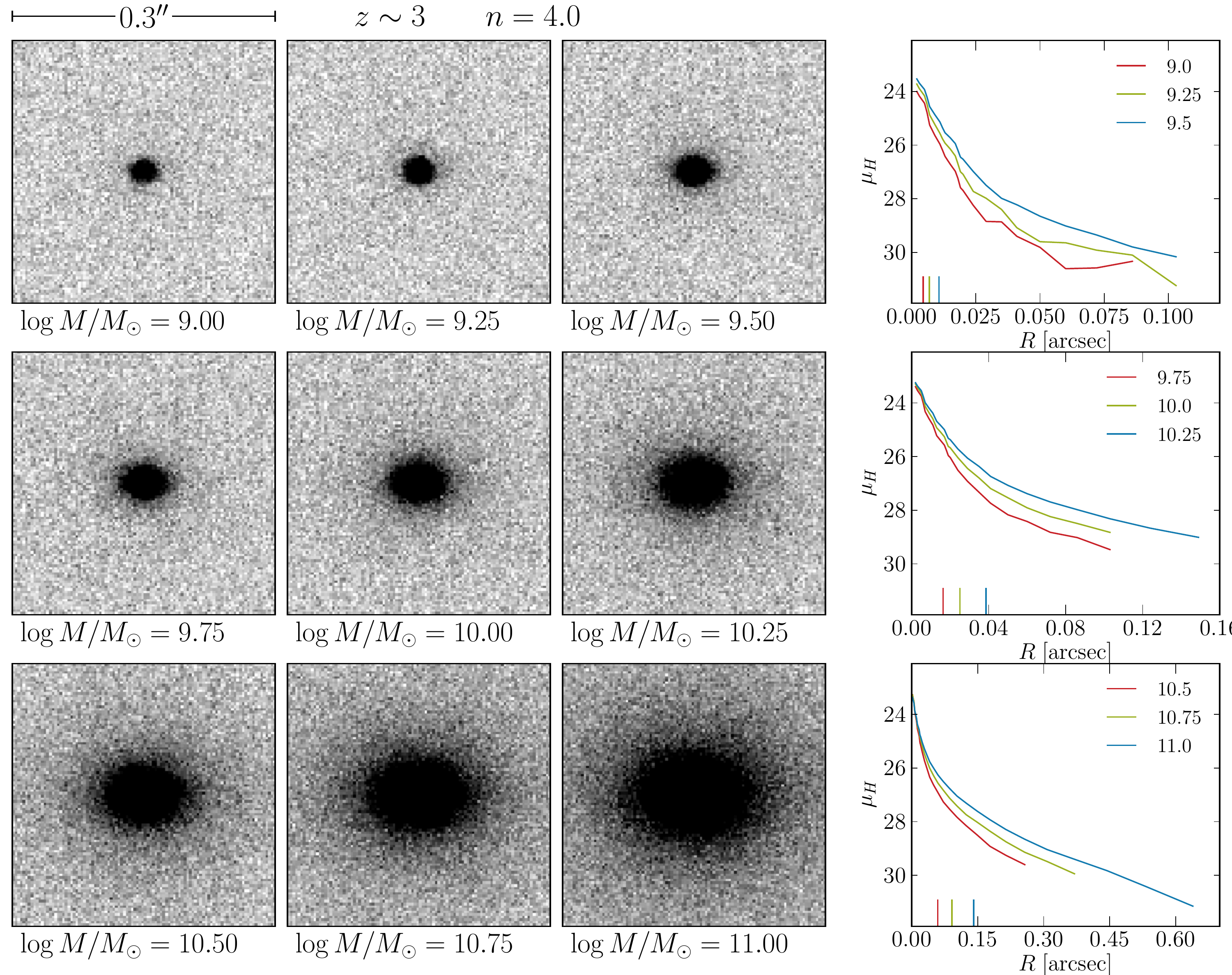}\\[1em]
  \includegraphics[width=\hsize]{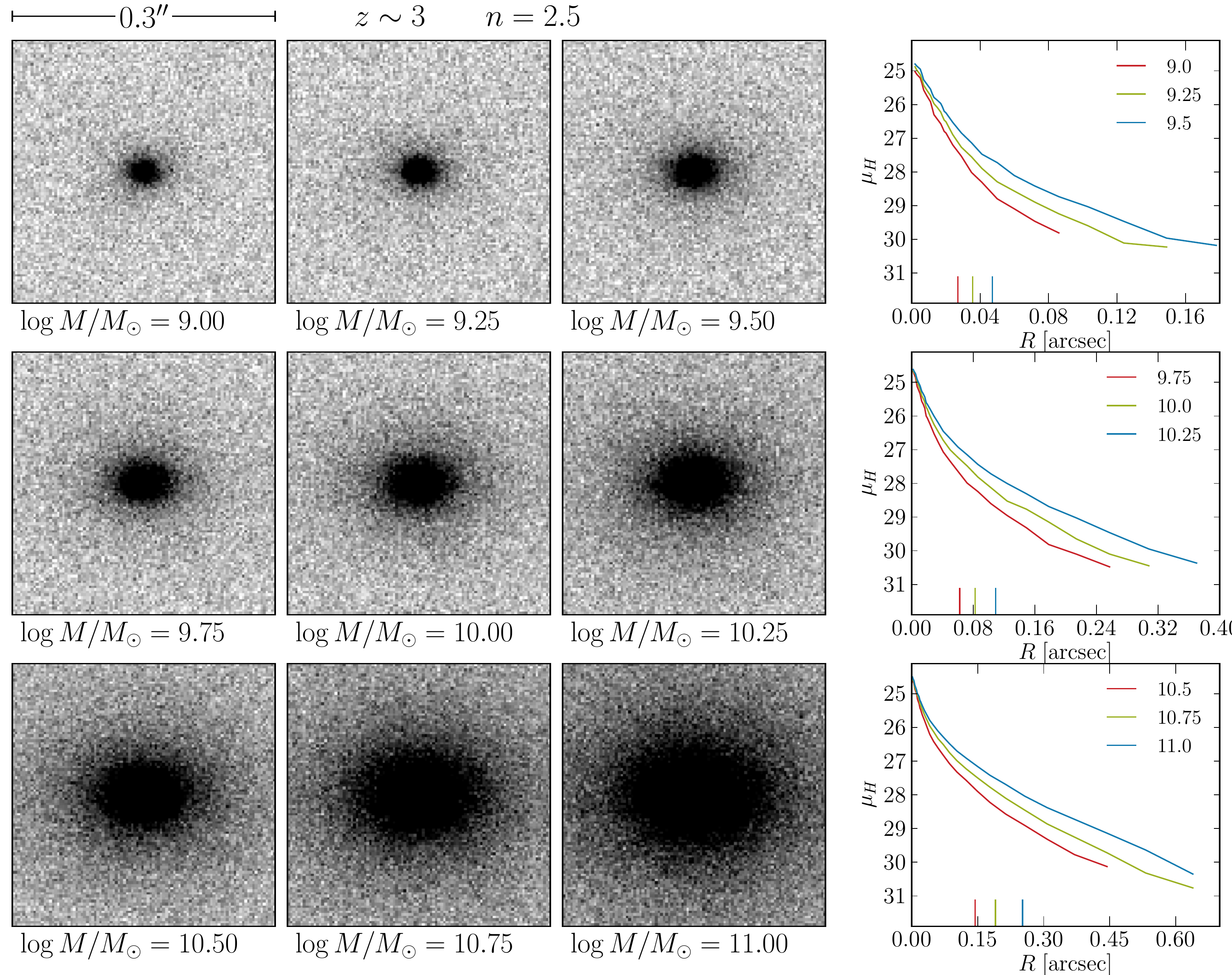}\\[1em]
  \includegraphics[width=\hsize]{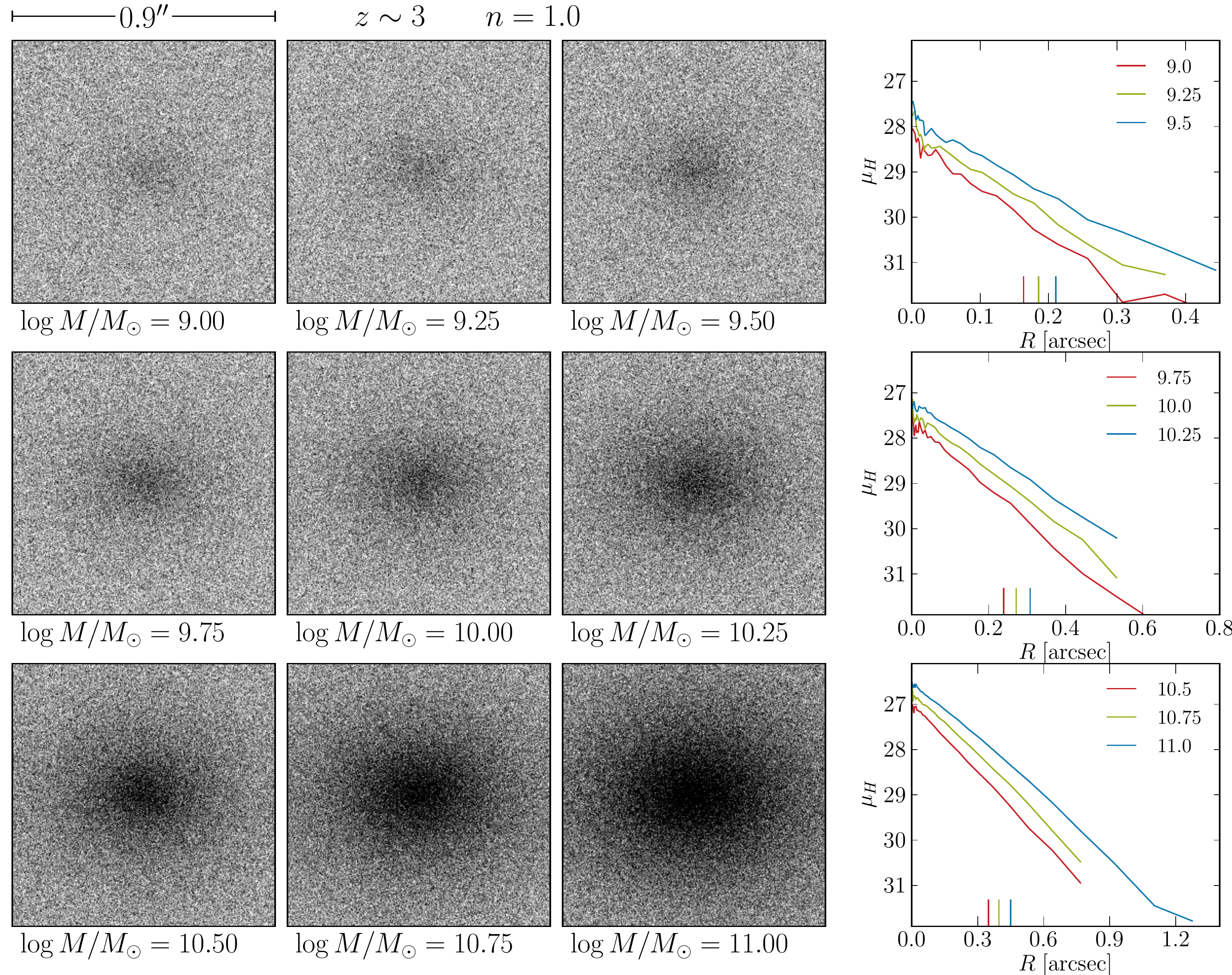}
  \caption{As in \ref{fig:imaz2_1} for $H$-band images of  $z=3$ galaxies.}
  \label{fig:imaHz3}
\end{figure}
\begin{figure}
  \centering
  \includegraphics[width=\hsize]{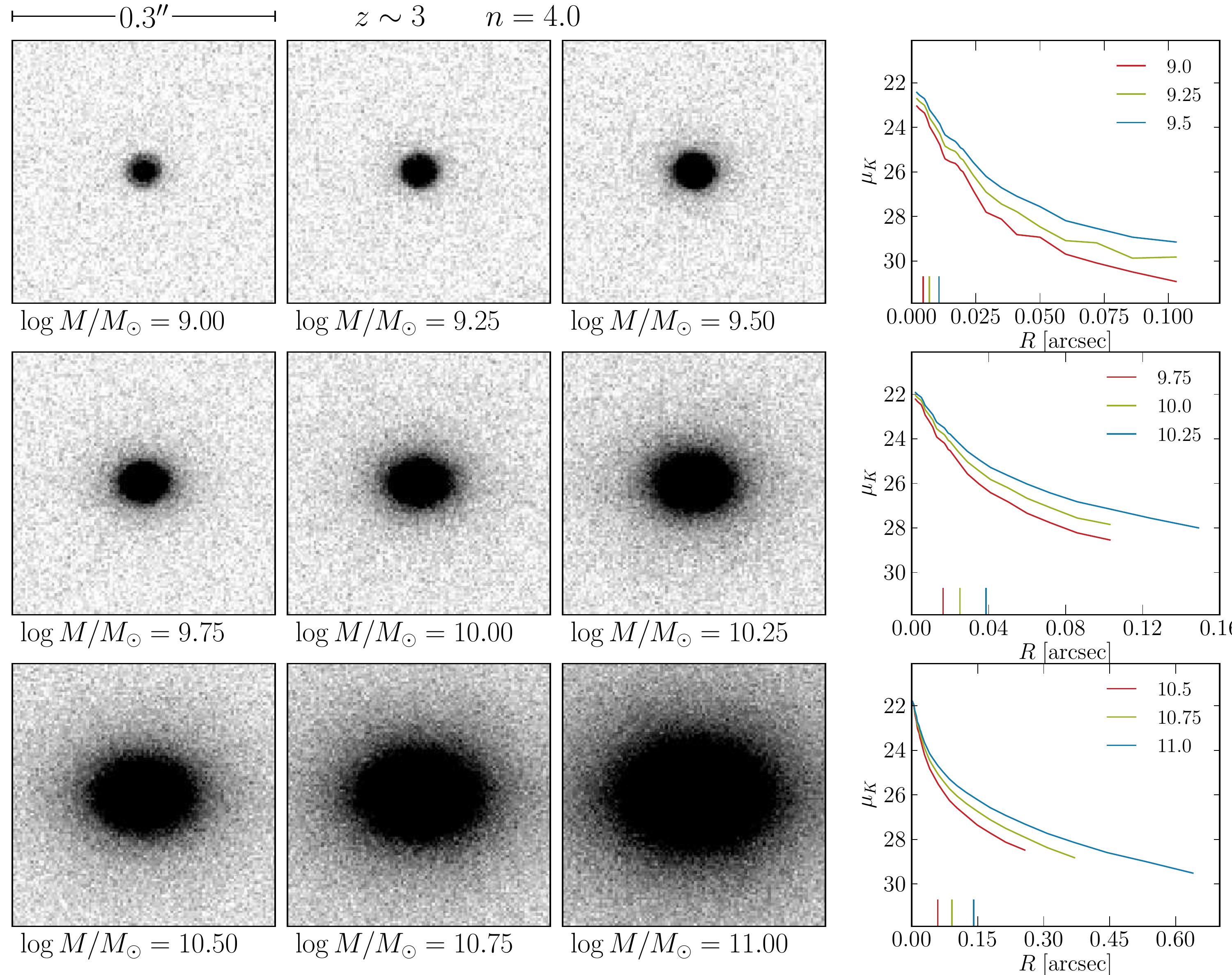}\\[1em]
  \includegraphics[width=\hsize]{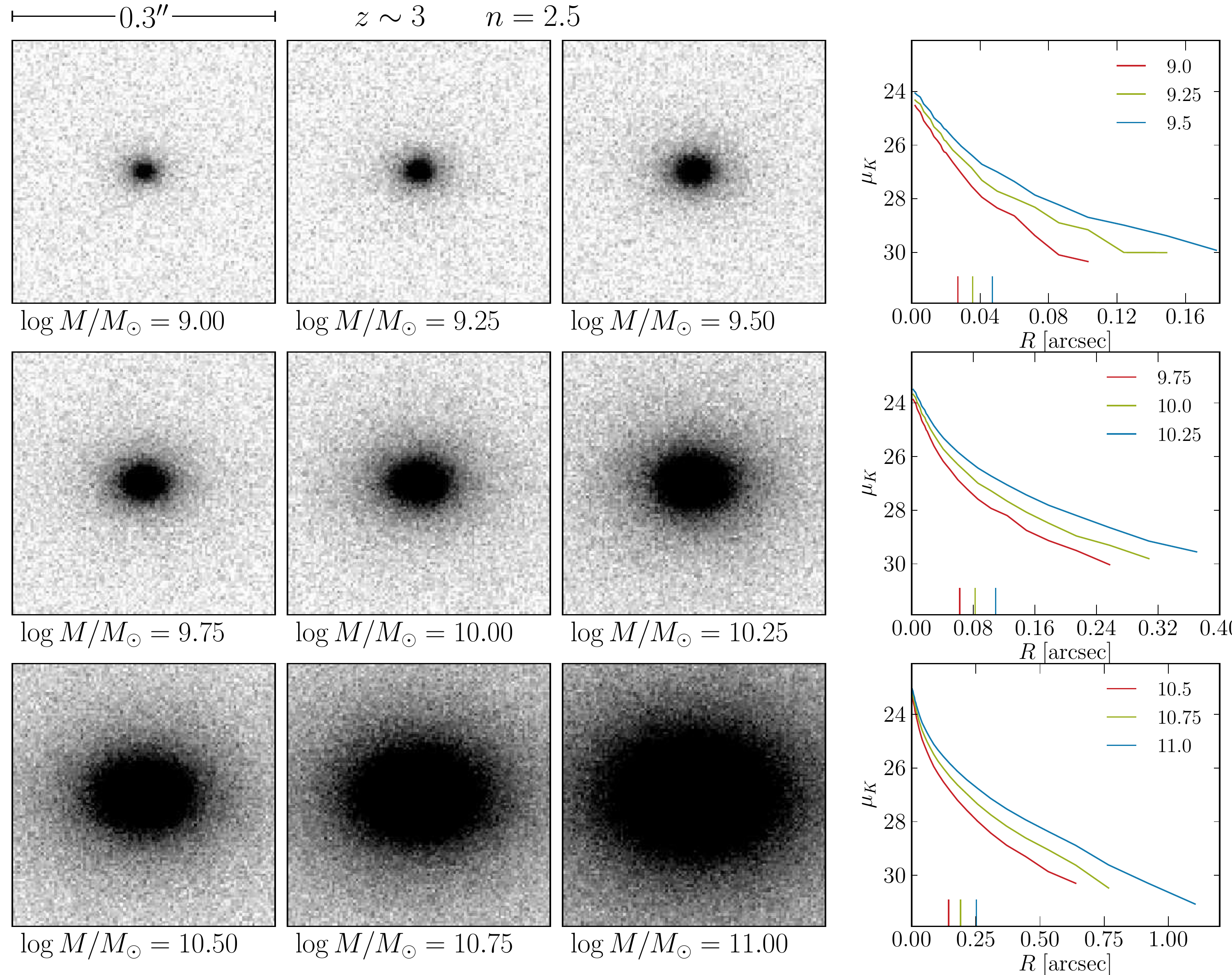}\\[1em]
  \includegraphics[width=\hsize]{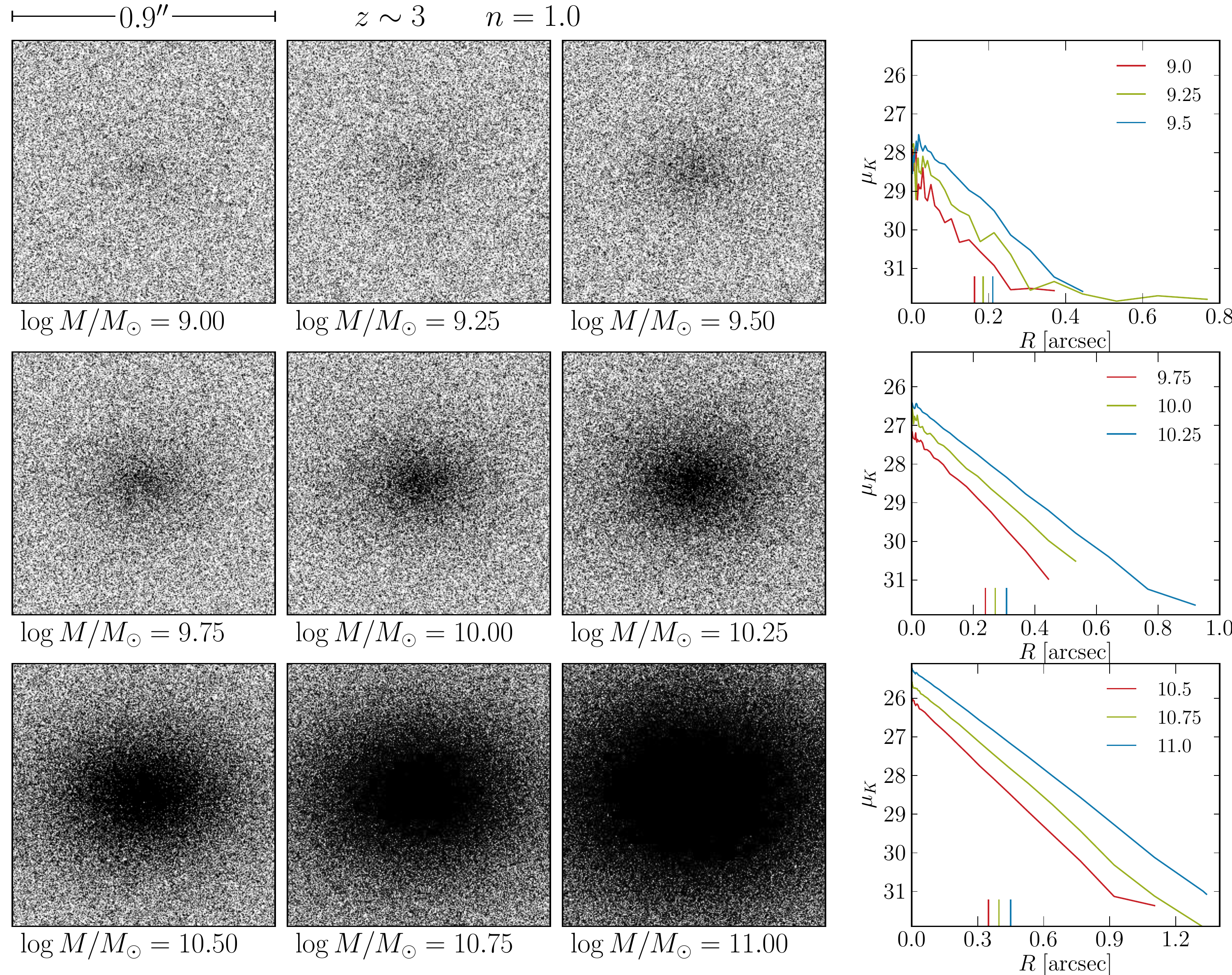}
  \caption{As in \ref{fig:imaz2_1} for $K$-band images of  $z=3$ galaxies.}
  \label{fig:imaKz3}
\end{figure}

\begin{table*}[!ht]
    \caption{Mean values and dispersion of the distribution of residuals in measured effective radius, Sersic index, and magnitude in $J$- and $H$-band simulations of $z=2$ galaxies.}
    \label{tab:paramsz2}
    \centering       
    \begin{tabular}{c r@{$\pm$}l r@{$\pm$}l r@{$\pm$}l r@{$\pm$}l r@{$\pm$}l r@{$\pm$}l }
      \hline\hline                 
      $\log M/M_\odot$&
      \multicolumn{2}{c}{$\Delta n/n^J$  [\%]}&
      \multicolumn{2}{c}{$\Delta R_e/R_e^J$ [\%]}  &
      \multicolumn{2}{c}{$\Delta J$}&
      \multicolumn{2}{c}{$\Delta n/n^H$  [\%]}&
      \multicolumn{2}{c}{$\Delta R_e/R_e^H$ [\%]} &
      \multicolumn{2}{c}{$\Delta H$}\\
      \hline
\multicolumn{13}{c}{$n=1.0$} \\
$ 9.00$& $  1.2$& $  5.5$& $  0.2$& $  2.0$& $ -0.00$& $  0.02$& $ -1.1$& $  6.2$& $ -0.3$& $  3.6$& $ -0.00$& $  0.03$\\
$ 9.25$& $  0.1$& $  3.3$& $  0.2$& $  1.8$& $ -0.01$& $  0.02$& $ -0.6$& $  4.5$& $  0.4$& $  2.2$& $  0.00$& $  0.02$\\
$ 9.50$& $ -0.2$& $  3.2$& $  0.4$& $  1.8$& $ -0.00$& $  0.01$& $ -0.3$& $  3.3$& $  0.1$& $  1.6$& $ -0.00$& $  0.02$\\
$ 9.75$& $ -0.3$& $  2.8$& $ -0.4$& $  1.2$& $  0.00$& $  0.01$& $ -1.0$& $  1.9$& $  0.1$& $  1.2$& $ -0.00$& $  0.01$\\
$10.00$& $ -0.7$& $  2.4$& $ -0.1$& $  1.4$& $  0.00$& $  0.01$& $ -0.6$& $  1.4$& $ -0.0$& $  0.9$& $  0.00$& $  0.01$\\
$10.25$& $  0.0$& $  1.8$& $  0.1$& $  0.9$& $ -0.00$& $  0.01$& $ -0.9$& $  1.3$& $ -0.2$& $  0.5$& $  0.00$& $  0.00$\\
$10.50$& $ -0.6$& $  1.7$& $ -0.2$& $  0.8$& $  0.00$& $  0.01$& $ -0.4$& $  1.0$& $  0.0$& $  0.5$& $  0.00$& $  0.01$\\
$10.75$& $  0.3$& $  1.4$& $  0.0$& $  0.7$& $  0.00$& $  0.01$& $ -0.3$& $  0.6$& $  0.0$& $  0.3$& $  0.00$& $  0.00$\\
$11.00$& $ -0.1$& $  1.3$& $  0.1$& $  0.7$& $  0.00$& $  0.01$& $ -0.4$& $  0.5$& $ -0.0$& $  0.2$& $  0.00$& $  0.00$\\
\multicolumn{13}{c}{$n=2.5$} \\
$ 9.00$& $  0.2$& $  4.4$& $ -1.9$& $  3.4$& $ -0.01$& $  0.02$& $  0.3$& $  3.0$& $  0.4$& $  3.3$& $ -0.01$& $  0.02$\\
$ 9.25$& $ -1.3$& $  4.4$& $ -2.5$& $  3.6$& $ -0.01$& $  0.02$& $ -0.6$& $  1.4$& $ -0.7$& $  2.0$& $ -0.01$& $  0.01$\\
$ 9.50$& $  0.4$& $  2.5$& $ -0.9$& $  2.7$& $ -0.01$& $  0.01$& $ -0.2$& $  1.6$& $ -0.3$& $  1.4$& $ -0.01$& $  0.01$\\
$ 9.75$& $  0.3$& $  3.2$& $ -1.2$& $  3.9$& $ -0.01$& $  0.02$& $ -0.3$& $  1.5$& $ -0.4$& $  1.6$& $ -0.01$& $  0.01$\\
$10.00$& $  0.3$& $  2.0$& $ -1.3$& $  2.8$& $ -0.00$& $  0.02$& $ -0.4$& $  1.1$& $ -0.3$& $  1.2$& $ -0.01$& $  0.01$\\
$10.25$& $  0.5$& $  2.3$& $ -1.0$& $  2.5$& $ -0.00$& $  0.01$& $ -0.2$& $  0.9$& $ -0.2$& $  1.3$& $ -0.01$& $  0.01$\\
$10.50$& $  0.4$& $  1.8$& $ -0.9$& $  1.8$& $  0.00$& $  0.01$& $ -0.5$& $  0.6$& $ -0.7$& $  1.0$& $ -0.00$& $  0.00$\\
$10.75$& $  0.1$& $  1.9$& $ -1.1$& $  1.8$& $ -0.00$& $  0.01$& $ -0.6$& $  0.4$& $ -1.1$& $  0.6$& $  0.00$& $  0.00$\\
$11.00$& $  0.2$& $  1.5$& $ -0.8$& $  1.6$& $  0.00$& $  0.01$& $ -0.7$& $  0.4$& $ -1.4$& $  0.5$& $  0.00$& $  0.00$\\
\multicolumn{13}{c}{$n=4.0$} \\
$ 9.00$& $-16.1$& $  8.2$& $ -3.7$& $  5.6$& $  0.05$& $  0.03$& $-14.9$& $  3.4$& $  4.6$& $  3.1$& $  0.03$& $  0.01$\\
$ 9.25$& $ -1.7$& $  9.5$& $  0.0$& $  6.2$& $  0.00$& $  0.03$& $ -9.3$& $  2.2$& $  1.5$& $  1.8$& $  0.01$& $  0.01$\\
$ 9.50$& $ -1.8$& $  5.3$& $ -1.5$& $  5.7$& $  0.00$& $  0.03$& $ -6.8$& $  2.0$& $ -0.5$& $  1.8$& $  0.01$& $  0.01$\\
$ 9.75$& $ -1.6$& $  4.3$& $ -2.8$& $  5.3$& $ -0.00$& $  0.02$& $ -4.0$& $  1.6$& $ -0.7$& $  1.9$& $  0.00$& $  0.01$\\
$10.00$& $  0.4$& $  4.2$& $ -1.6$& $  4.5$& $ -0.01$& $  0.02$& $ -2.8$& $  1.4$& $ -1.5$& $  2.0$& $  0.00$& $  0.01$\\
$10.25$& $ -0.7$& $  3.1$& $ -2.7$& $  3.9$& $ -0.00$& $  0.02$& $ -2.3$& $  1.1$& $ -1.7$& $  1.8$& $  0.00$& $  0.01$\\
$10.50$& $  1.0$& $  2.7$& $ -2.1$& $  4.5$& $ -0.00$& $  0.02$& $ -1.1$& $  0.8$& $ -1.1$& $  1.7$& $ -0.00$& $  0.01$\\
$10.75$& $  0.2$& $  1.6$& $ -2.3$& $  3.4$& $  0.00$& $  0.02$& $ -0.8$& $  0.8$& $ -1.1$& $  1.8$& $ -0.00$& $  0.01$\\
$11.00$& $  0.5$& $  1.8$& $ -2.0$& $  2.7$& $  0.00$& $  0.02$& $ -0.9$& $  0.4$& $ -2.0$& $  1.1$& $  0.00$& $  0.01$\\
      \hline                                             
    \end{tabular}
  \end{table*}

\begin{table*}[!ht]
    \caption{Mean values and dispersion of the distribution of residuals in measured effective radius, Sersic index, and magnitude in $J$- and $H$-band simulations of $z=3$ galaxies.}
    \label{tab:paramsz3}
    \centering       
    \begin{tabular}{c r@{$\pm$}l r@{$\pm$}l r@{$\pm$}l r@{$\pm$}l r@{$\pm$}l r@{$\pm$}l }
      \hline\hline                 
      $\log M/M_\odot$&
      \multicolumn{2}{c}{$\Delta n/n^H$  [\%]}&
      \multicolumn{2}{c}{$\Delta R_e/R_e^H$ [\%]}  &
      \multicolumn{2}{c}{$\Delta H$}&
      \multicolumn{2}{c}{$\Delta n/n^K$  [\%]}&
      \multicolumn{2}{c}{$\Delta R_e/R_e^K$ [\%]} &
      \multicolumn{2}{c}{$\Delta K$}\\
      \hline
\multicolumn{13}{c}{$n=1.0$} \\
$ 9.00$& $ -0.2$& $  8.9$& $  1.1$& $  4.5$& $ -0.02$& $  0.04$& $ -6.8$& $ 16.4$& $ -5.4$& $  9.3$& $  0.06$& $  0.08$\\
$ 9.25$& $ -3.7$& $  5.3$& $ -1.7$& $  3.6$& $  0.01$& $  0.03$& $  1.0$& $  9.2$& $ -0.0$& $  6.5$& $  0.01$& $  0.06$\\
$ 9.50$& $  0.4$& $  3.7$& $  0.4$& $  2.4$& $ -0.01$& $  0.02$& $  0.2$& $  5.0$& $  2.1$& $  3.7$& $ -0.00$& $  0.04$\\
$ 9.75$& $ -1.8$& $  3.0$& $ -0.4$& $  1.8$& $  0.00$& $  0.02$& $ -1.2$& $  3.2$& $  0.8$& $  2.9$& $  0.01$& $  0.02$\\
$10.00$& $ -1.8$& $  3.2$& $ -0.4$& $  1.4$& $  0.00$& $  0.01$& $  0.8$& $  1.9$& $  1.6$& $  1.4$& $  0.01$& $  0.01$\\
$10.25$& $ -1.4$& $  1.7$& $ -0.2$& $  0.9$& $  0.00$& $  0.01$& $  1.5$& $  1.5$& $  1.9$& $  0.9$& $  0.00$& $  0.01$\\
$10.50$& $ -0.5$& $  1.1$& $  0.1$& $  0.7$& $  0.00$& $  0.01$& $  1.4$& $  0.7$& $  1.7$& $  0.5$& $  0.00$& $  0.01$\\
$10.75$& $ -0.3$& $  0.8$& $ -0.1$& $  0.6$& $  0.00$& $  0.00$& $  1.2$& $  0.6$& $  1.8$& $  0.4$& $ -0.00$& $  0.00$\\
$11.00$& $ -0.2$& $  0.7$& $  0.1$& $  0.4$& $  0.00$& $  0.00$& $  1.1$& $  0.3$& $  1.7$& $  0.3$& $ -0.00$& $  0.00$\\
\multicolumn{13}{c}{$n=2.5$} \\
$ 9.00$& $ -1.3$& $  5.4$& $  0.5$& $  5.1$& $ -0.00$& $  0.03$& $ -1.4$& $  4.9$& $  0.3$& $  4.6$& $  0.02$& $  0.03$\\
$ 9.25$& $ -0.7$& $  3.6$& $  0.0$& $  3.9$& $ -0.01$& $  0.02$& $ -1.5$& $  3.5$& $ -1.1$& $  3.2$& $  0.02$& $  0.02$\\
$ 9.50$& $  0.0$& $  2.2$& $ -0.1$& $  2.3$& $ -0.01$& $  0.01$& $  0.1$& $  1.7$& $ -0.2$& $  2.2$& $  0.01$& $  0.01$\\
$ 9.75$& $ -0.1$& $  1.8$& $  0.1$& $  1.9$& $ -0.01$& $  0.01$& $ -0.3$& $  1.6$& $  0.1$& $  1.9$& $  0.02$& $  0.01$\\
$10.00$& $ -0.2$& $  1.6$& $  0.3$& $  1.8$& $ -0.01$& $  0.01$& $ -0.1$& $  0.9$& $ -0.4$& $  1.2$& $  0.02$& $  0.01$\\
$10.25$& $ -0.5$& $  1.3$& $ -0.4$& $  1.4$& $ -0.01$& $  0.01$& $ -0.1$& $  0.7$& $  0.3$& $  1.1$& $  0.01$& $  0.01$\\
$10.50$& $ -0.6$& $  0.9$& $ -0.6$& $  1.2$& $ -0.00$& $  0.01$& $  0.4$& $  0.6$& $  0.8$& $  0.9$& $  0.01$& $  0.00$\\
$10.75$& $ -0.6$& $  0.7$& $ -0.8$& $  1.1$& $ -0.00$& $  0.01$& $  0.7$& $  0.3$& $  1.2$& $  0.7$& $  0.01$& $  0.00$\\
$11.00$& $ -0.6$& $  0.6$& $ -1.1$& $  0.9$& $  0.00$& $  0.01$& $  1.0$& $  0.3$& $  1.7$& $  0.5$& $  0.00$& $  0.00$\\
\multicolumn{13}{c}{$n=4.0$} \\
$ 9.00$& $-17.6$& $  6.2$& $  7.8$& $  3.9$& $  0.04$& $  0.02$& $-12.1$& $  4.8$& $  7.8$& $  2.3$& $  0.06$& $  0.01$\\
$ 9.25$& $-11.7$& $  4.2$& $  2.2$& $  3.1$& $  0.03$& $  0.02$& $ -8.8$& $  2.5$& $  2.9$& $  1.3$& $  0.05$& $  0.01$\\
$ 9.50$& $ -8.2$& $  2.6$& $ -0.2$& $  2.6$& $  0.01$& $  0.01$& $ -6.3$& $  2.1$& $  0.7$& $  1.5$& $  0.03$& $  0.01$\\
$ 9.75$& $ -6.2$& $  2.1$& $ -1.5$& $  2.4$& $  0.01$& $  0.01$& $ -3.8$& $  1.3$& $ -0.1$& $  1.4$& $  0.03$& $  0.01$\\
$10.00$& $ -3.8$& $  1.8$& $ -1.7$& $  2.2$& $  0.00$& $  0.01$& $ -2.7$& $  0.9$& $ -0.6$& $  1.3$& $  0.02$& $  0.01$\\
$10.25$& $ -2.1$& $  1.5$& $ -1.1$& $  2.0$& $  0.00$& $  0.01$& $ -1.8$& $  0.8$& $ -1.1$& $  1.2$& $  0.02$& $  0.00$\\
$10.50$& $ -1.4$& $  1.0$& $ -1.2$& $  1.9$& $ -0.00$& $  0.01$& $ -1.0$& $  0.4$& $ -0.5$& $  1.0$& $  0.02$& $  0.00$\\
$10.75$& $ -1.0$& $  0.6$& $ -1.1$& $  1.3$& $ -0.00$& $  0.01$& $ -0.6$& $  0.5$& $ -0.4$& $  1.4$& $  0.02$& $  0.00$\\
$11.00$& $ -0.9$& $  0.6$& $ -1.6$& $  1.4$& $ -0.00$& $  0.01$& $  0.1$& $  0.3$& $  0.6$& $  1.2$& $  0.01$& $  0.00$\\
      \hline                                             
    \end{tabular}
  \end{table*}

\begin{table}
    \caption{Mean values and dispersion of the measured colour gradients measured in  $z=2$ simulated galaxies galaxies.}
    \label{tab:gradsz2}
    \centering       
    \begin{tabular}{c r@{$\pm$}l r@{$\pm$}l r@{$\pm$}l }
      \hline\hline  
      $\log M/M_\odot$&
      \multicolumn{6}{c}{$\nabla(U-V)_\text{restframe}$}\\
      &
      \multicolumn{2}{c}{$n=1.0$}&
      \multicolumn{2}{c}{$n=2.5$}&
      \multicolumn{2}{c}{$n=4.0$}\\
      &\multicolumn{2}{c}{mag/dex}&\multicolumn{2}{c}{mag/dex}&\multicolumn{2}{c}{mag/dex}\\
      \hline
$  9.00$ & $  0.016$ & $  0.119$ & $  0.018$ & $  0.060$ & $  0.061$ & $  0.133$\\
$  9.25$ & $  0.021$ & $  0.092$ & $  0.018$ & $  0.047$ & $  0.104$ & $  0.089$\\
$  9.50$ & $  0.004$ & $  0.074$ & $  0.022$ & $  0.037$ & $  0.067$ & $  0.039$\\
$  9.75$ & $  0.007$ & $  0.058$ & $  0.009$ & $  0.029$ & $  0.052$ & $  0.032$\\
$ 10.00$ & $ -0.002$ & $  0.042$ & $  0.022$ & $  0.026$ & $  0.041$ & $  0.023$\\
$ 10.25$ & $  0.009$ & $  0.033$ & $  0.020$ & $  0.025$ & $  0.028$ & $  0.028$\\
$ 10.50$ & $ -0.003$ & $  0.031$ & $  0.014$ & $  0.025$ & $  0.030$ & $  0.021$\\
$ 10.75$ & $  0.004$ & $  0.022$ & $  0.009$ & $  0.020$ & $  0.020$ & $  0.021$\\
$ 11.00$ & $  0.001$ & $  0.022$ & $  0.006$ & $  0.021$ & $  0.013$ & $  0.019$\\
      \hline        

    \end{tabular}
  \end{table}
  
\begin{table}
    \caption{Mean values and dispersion of the measured colour gradients measured in  $z=3$ simulated galaxies galaxies.}
    \label{tab:gradsz3}
    \centering       
    \begin{tabular}{c r@{$\pm$}l r@{$\pm$}l r@{$\pm$}l }
      \hline\hline  
      $\log M/M_\odot$&
      \multicolumn{6}{c}{$\nabla(U-V)_\text{restframe}$}\\
      &
      \multicolumn{2}{c}{$n=1.0$}&
      \multicolumn{2}{c}{$n=2.5$}&
      \multicolumn{2}{c}{$n=4.0$}\\
      &\multicolumn{2}{c}{mag/dex}&\multicolumn{2}{c}{mag/dex}&\multicolumn{2}{c}{mag/dex}\\
      \hline
$  9.00$ & $  0.019$ & $  0.355$ & $ -0.003$ & $  0.067$ & $ -0.079$ & $  0.102$\\
$  9.25$ & $ -0.076$ & $  0.166$ & $  0.005$ & $  0.049$ & $ -0.028$ & $  0.052$\\
$  9.50$ & $  0.025$ & $  0.113$ & $ -0.002$ & $  0.033$ & $ -0.017$ & $  0.029$\\
$  9.75$ & $  0.004$ & $  0.064$ & $  0.001$ & $  0.025$ & $ -0.013$ & $  0.017$\\
$ 10.00$ & $ -0.015$ & $  0.054$ & $ -0.008$ & $  0.018$ & $ -0.005$ & $  0.013$\\
$ 10.25$ & $ -0.014$ & $  0.036$ & $ -0.000$ & $  0.015$ & $ -0.003$ & $  0.009$\\
$ 10.50$ & $ -0.008$ & $  0.019$ & $  0.003$ & $  0.013$ & $ -0.000$ & $  0.011$\\
$ 10.75$ & $  0.001$ & $  0.015$ & $  0.005$ & $  0.012$ & $  0.003$ & $  0.013$\\
$ 11.00$ & $  0.004$ & $  0.011$ & $  0.010$ & $  0.011$ & $  0.008$ & $  0.012$\\
      \hline        

    \end{tabular}
  \end{table}

\end{appendix}
\end{document}